\documentclass{aa} 
\usepackage{graphicx}
\usepackage{lscape}
\usepackage{subfigure}
\usepackage{natbib}
\usepackage[varg]{txfonts}
\usepackage{longtable}

\bibpunct{(}{)}{;}{a}{}{,} 
\begin{document}

   \title{Variability-selected active galactic nuclei in the \emph{VST}-SUDARE/VOICE survey of the COSMOS field\thanks{Observations were provided by the ESO programs 088.D-0370 and 088.D-4013 (PI G. Pignata)}}

   \author{D. De Cicco \inst{1}, M. Paolillo\inst{1,2,3}, G. Covone\inst{1,2}, S. Falocco\inst{1,2}, G. Longo\inst{1,4}, A. Grado\inst{5}, L. Limatola\inst{5}, M. T. Botticella\inst{5}, G. Pignata\inst{6,7}, E. Cappellaro\inst{8}, M. Vaccari\inst{9}, D. Trevese\inst{10}, F. Vagnetti\inst{11}, M. Salvato\inst{12}, M. Radovich\inst{8},\\W. N. Brandt\inst{13,14}, M. Capaccioli\inst{1}, N. R. Napolitano\inst{5}, P. Schipani\inst{5}}
   \titlerunning{Optical variability-selected AGNs in the COSMOS field}
   \authorrunning{D. De Cicco et al.}

  \institute{Department of Physics, University of Napoli ``Federico II'', via Cinthia 9, 80126 Napoli, Italy\\e-mail: demetra.decicco@unina.it
         \and
      		INFN - Sezione di Napoli, via Cinthia 9, 80126 Napoli, Italy 
	\and
		ASI Science Data Center, via del Politecnico snc, 00133 Roma, Italy 
	\and
		Visiting associate - Department of Astronomy, California Institute of Technology, CA 90125, USA 
	\and
		INAF - Osservatorio Astronomico di Capodimonte, via Moiariello 16, 80131 Napoli, Italy 
	\and
		Departamento de Ciencias Fisicas, Universidad Andres Bello, Avda. Republica 252, Santiago, Chile 
	\and
		Millennium Institute of Astrophysics, Santiago, Chile 
	\and
		INAF - Osservatorio Astronomico di Padova, vicolo dell'Osservatorio 5, I-35122 Padova, Italy 
	\and
	 	Astrophysics Group, Department of Physics, University of the Western Cape, Private Bag X17, 7535 Bellville, Cape Town, South Africa 
	\and
		Department of Physics, University of Roma ``La Sapienza'', Piazzale Aldo Moro 2, 00185 Roma, Italy 
	\and
		Department of Physics, University of Roma ``Tor Vergata'', via della Ricerca Scientifica 1, 00133 Roma, Italy 
	\and
		Max Planck Institut f\"{u}r Extraterrestrische Physik, Giessenbachstra\ss e 1, D-85748 Garching bei M\"{u}nchen, Germany 
	\and
		Department of Astronomy and Astrophysics, The Pennsylvania State University, University Park, PA 16802, USA 
	\and
		Institute for Gravitation and the Cosmos, The Pennsylvania State University, University Park, PA 16802, USA\\}
   \date{}
  \abstract
   {Active galaxies are characterized by variability at every wavelength, with timescales from hours to years depending on the observing window. Optical variability has proven to be an effective way of detecting AGNs in imaging surveys, lasting from weeks to years.}
   {In the present work we test the use of optical variability as a tool to identify active galactic nuclei in the \emph{VST} multiepoch survey of the COSMOS field, originally tailored to detect supernova events.}
   {We make use of the multiwavelength data provided by other COSMOS surveys to discuss the reliability of the method and the nature of our AGN candidates.}
   {The selection on the basis of optical variability returns a sample of 83 AGN candidates; based on a number of diagnostics, we conclude that 67 of them are confirmed AGNs (81\% purity), 12 are classified as supernovae, while the nature of the remaining 4 is unknown. For the subsample of AGNs with some spectroscopic classification, we find that Type 1 are prevalent (89\%) compared to Type 2 AGNs (11\%).
   Overall, our approach is able to retrieve on average 15\% of all AGNs in the field identified by means of spectroscopic or X-ray classification, with a strong dependence on the source apparent magnitude (completeness ranging from $26\%$ to $5\%$). In particular, the completeness for Type 1 AGNs is 25\%, while it drops to 6\% for Type 2 AGNs. The rest of the X-ray selected AGN population presents on average a larger r.m.s. variability than the bulk of non-variable sources, indicating that variability detection for at least some of these objects is prevented only by the photometric accuracy of the data. The low completeness is in part due to the short observing span: we show that increasing the temporal baseline results in larger samples as expected for sources with a red-noise power spectrum. Our results allow us to assess the usefulness of this AGN selection technique in view of future wide-field surveys.}
   {}

   \keywords{galaxies: active -- X-rays: galaxies -- quasars: general -- supernovae: general -- surveys 
               }

   \maketitle
   
%

\section{Introduction}
It is now widely accepted that the engine powering an active galactic nucleus (AGN) is an accreting supermassive black hole (SMBH) at the center of the active galaxy. Several empirical relations connect some of the properties and physical quantities of the central black hole and the galaxy: e.g., correlation between black hole mass and stellar velocity dispersion \citep[e.g.,][]{Ferrarese&Merritt}), or between black hole mass and galaxy luminosity \citep[e.g.,][]{Kormendy&Ho,Kormendy&Richstone}; furthermore, evidence exists for a co-evolution of the quasi-stellar object (QSO) luminosity function and the star formation rate (SFR) with cosmic time \citep[e.g.,][]{Fiore}. Such relations support the existence of a tight feedback between SMBH and galaxy evolution. 
Since most massive galaxies host a SMBH, an extensive knowledge of the black hole demography is of primary importance to increase our understanding of galaxy evolution.

Several methods have been developed to identify AGNs; a single identification technique is generally not sufficient for a complete and bias-free census of the AGN population. X-ray emission is at present the most effective instrument for AGN identification: AGN spectra are broadband and are characterized by a considerable X-ray component, which is generally comparable to optical emission as regards its spectral extent and, above a certain luminosity, constitutes unequivocal evidence of the active nature of a galaxy \citep[e.g.,][]{Brandt&Hasinger}. Remarkable advantages of using X-ray emission to find AGNs are the high penetrating power, allowing us to detect even those AGNs that are obscured at other wavelengths, and the large-amplitude and fast variability. 
However, the space observations that are required to detect X-ray radiation have a higher cost and a more limited field of view (FoV) than ground-based observations.

The spectra of most AGNs are also characterized by prominent emission lines, broader than those found in the spectra of normal galaxies. Several diagnostics, e.g., the BPT diagram \citep{BPT}, allow us to identify these types of AGNs by means of the properties of their emission lines, provided that obscuration in the wavelength range of interest is not signficant; unfortunately, spectroscopy is a time-consuming technique for AGN identification, especially when dealing with very large samples and faint sources.

Provided that multicolor data are available, color selection is a widespread technique to find AGNs at UV/optical/IR wavelengths \citep[e.g.,][]{Fan}. 
Given their different spectral energy distributions (SEDs), the amount of light in the UV and IR bands is much higher for AGNs (unobscured, in the first case) than for stars or non-active galaxies, hence the corresponding flux ratios will be different and a color-color diagram will reveal the nature of different sources depending on their position in the plot \citep[e.g.,][]{Richards}. The technique can be refined by making use of a multidimensional color space; it is widely used as a few images suffice to get information about a large number of candidates. Nonetheless, several biases affect the method: first, the non-stellar nature of their color can only be used to identify those AGNs that are bright enough to outshine the host galaxy, while it does not work with faint\footnote{Faint AGNs are characterized by a nuclear absolute magnitude in the $r$ band $M_R \gtrsim -21.5$ mag \citep{Boutsia}.} AGNs because emission from the host galaxy dominates, swamping the nuclear light. 
In general, the classification of objects by means of color selection criteria needs great accuracy in order to minimize the contamination by stars. One more difficulty in AGN selection through color is absorption, which is attributable to the presence of dust in the plane of the Galaxy (affecting low-latitude observations) and also to extinction, intrinsic to the AGN itself or the host galaxy, whose importance increases with redshift \citep[e.g.,][]{Krolik}. 

Variability is a defining feature of AGN emission at all wavebands. Luminosity variations generally affect both continuum and broad-line emission; the timescales range from hours to years, depending on the observing wavelength \citep[e.g.,][]{Ulrich, Gaskell&Klimek}. Currently, variability is generally attributed to instabilities in the AGN accretion disk \citep[e.g.,][]{Pereyra}, possibly associated with other phenomena, such as changes in the accretion rate, presence of obscuring medium, star disruption, and gravitational microlensing \citep[e.g.,][]{Aretxaga&Terlevich}. Variability measurements can help in understanding the underlying emission mechanism, constraining the size and structure of the emitting region. Most QSOs typically exhibit continuum variations on the order of $10\%$ over months to years, while blazars show even more substantial variations on much shorter timescales (sometimes even on the order of minutes; see, e.g., \citealt{Albert}). The extent of variations is not the same at all wavelength ranges. Optical continuum variability seems to be a universal feature of broad-line AGNs (BLAGNs) on timescales from months to years \citep[e.g.,][]{Webb&Malkan}, with variations ranging from $10^{-2}$ to $10^{-1}$ mag, while a magnitude variation around $30\%$ is common in the X-ray region \citep{Paolillo}. 

It is not yet clear whether optical variability is an intrinsic phenomenon, or if it originates in the reprocessing of X-ray emission from the inner regions of the disk; several models have been developed, attempting to investigate a possible relation between optical/UV and X-ray variability. One of the most widespread theories explains the first as a consequence of the reprocessing of the second: according to this interpretation, X-ray radiation interacts with the surrounding, cooler matter (disk, torus), thus losing energy \citep[e.g.,][]{Krolik}. Conversely, other models suggest that X-ray emission follows from optical/UV radiation, because of the Comptonization of optical/UV photons by relativistic electrons. Since results from different observing campaigns are conflicting (see, e.g., \citealt{Uttley}), no theory dominates at present and it is likely that both processes occur at the same time.

Optical variability has been widely used to identify unobscured AGNs in multiepoch surveys \citep[e.g.,][]{Bershady, Klesman&Sarajedini, Trevese, Villforth1}; the techniques based on optical variability allow the surveying of extended areas by means of ground-based telescopes and do not miss those sources that are characterized by an unusually low X-ray to optical flux ratio (X/O) and hence are not detected by X-rays surveys; furthermore, as several studies \citep[e.g.,][]{Barr&Mushotzky, Lawrence&Papadakis, Cristiani} support the existence of an anti-correlation between AGN luminosity and the amplitude of variability, low-luminosity AGNs (LLAGNs) can be identified more effectively. 

The present work aims at detecting AGNs in the COSMOS field on the basis of their optical variability, and at verifying the effectiveness of this method against other traditional approaches, taking advantage of the extensive multiwavelength coverage. The AGN search was conducted exploiting the data acquired by the COSMOS extension of the SUDARE supernova survey program (see Section \ref{section:dataset}) by the \emph{VLT} Survey Telescope (\emph{VST}); with its five-month baseline, the program is a suitable tool for the selection of variable sources in the optical wavelength range, given the typical timescales on the order of days. The COSMOS field is one of the most widely surveyed areas in the sky; data from multiwavelength surveys are used to confirm the nature of our sample of AGN candidates.

The paper is organized as follows: in Section \ref{section:dataset} we introduce the telescope and the survey and describe our dataset; Section \ref{section:selection} illustrates the procedure used to select our sample of optically variable sources and how we refine it in order to get a reliable AGN candidate sample; in Section \ref{section:validation} we deal with the other COSMOS catalogs and the diagnostics we made use of to confirm the nature of the sources in our sample; and in Section \ref{section:discussion} we discuss our findings and final results.

\section{The SUDARE survey with the \emph{VST}}
\label{section:dataset}
The \emph{VLT} Survey Telescope (\emph{VST}; \citealt{VST}) is located at Cerro Paranal Observatory; it is a joint venture between the European Southern Observatory (ESO) and the INAF-Osservatorio Astronomico di Capodimonte (OAC) in Napoli.
The telescope is 2.65 m in diameter and is equipped with the single focal plane detector OmegaCAM \citep{Kuijken}: a mosaic of 32 CCD detectors made up of 268 megapixels in total, corresponding to a 26 cm $\times$ 26 cm area and a $1^\circ \times 1^\circ$ FoV, the resolution being $0.214\arcsec$/pixel.
The \emph{VST} is dedicated to surveys in the wavelength range 0.3 -- 1.0 $\mu$m. 

At present the \emph{VST} is involved in many Galactic and extragalactic survey programs; the present work is based on the analysis of images from the SUpernova Diversity And Rate Evolution (SUDARE) survey \citep{Botticella}. 
SUDARE is a supernova search program designed to measure the rates of the different supernova (SN) types at medium redshifts ($0.3<z<0.8$) and investigate their correlation with the properties of the host galaxies.
SUDARE observations cover two different sky regions, centered on the COSMOS field (named after the Cosmic Evolution Survey; see \citealt{Scoville1}) and the \emph{Chandra} Deep Field South (CDFS). The CDFS field is also covered by the \emph{VST} Optical Imaging of the CDFS and ES1 (VOICE; \citealt{Vaccari}) survey, providing deep \emph{ugri} stacks over a 2 square degree area centered on the CDFS. Our analysis concerns the COSMOS field; a complementary study focusing on the CDFS region will be presented in \citet{Falocco}.

The survey provides data in the \textit{g}, \textit{r}, and \textit{i} bands, with an observing frequency of approximately ten days for the \textit{g} and \textit{i} bands and three days for the \textit{r} band, depending on the various observational constraints.
We discuss here the analysis of 28 epochs in the \textit{r} band, for which we have the best temporal sampling. The observations cover the period from December 2011 to May 2012. Each epoch is made up of five dithered exposures of the field, for a total exposure time of 1800 s per epoch\footnote{Throughout the present paper, the word ``epoch'' will always refer to the combination of five images -- hereafter exposures -- corresponding to the same observing block (OB).}. 
We also obtained a stacked image, combining all the exposures (55 in total) with a seeing FWHM $< 0.8$\arcsec, corresponding to a total exposure time of 19800 s. The limiting magnitude of the stacked image at $\sim 5\sigma$ above the background r.m.s. is \emph{r}(AB) $\approx26$ mag, while for single epochs it is generally \emph{r}(AB) $\lesssim 25$ mag. The chosen parameters for source extraction are not the best to take advantage of the full depth of the survey, but are suitable to our variability analysis, which is focused on rather bright sources (\emph{r}(AB) $< 23$ mag; see below); the full catalogs will be presented in forthcoming papers \citep{Vaccari}.
Throughout our analysis we used the coordinates of the sources detected in the stacked image as reference coordinates for catalog matching.
Table \ref{tab:epochs} gives the names, dates, and seeing FWHM of the individual epochs and the stacked image used in our variability analysis. 
We excluded one epoch (corresponding to the OB 611379) since the exposures are strongly affected by aesthetic artifacts.
The data regarding the excluded epoch are given in Table \ref{tab:epochs} as well.

\begin{table}[tbp]
\caption{COSMOS dataset}             
\label{tab:epochs}      
\centering
\begin{tabular}{c c c c}
\hline\hline
epoch & OB-ID & obs. date & seeing (FWHM)\\
 & & & (arcsec)\\
\hline
1 & $\mbox{611279}$ & 2011-Dec-18 & $0.64 $\\
2 & $\mbox{611283}$ & 2011-Dec-22 & $0.94 $\\
3 & $\mbox{611287}$ & 2011-Dec-27 & $1.04 $\\
4 & $\mbox{611291}$ & 2011-Dec-31 & $1.15 $\\
5 & $\mbox{611295}$ & 2012-Jan-02 & $0.67 $\\
6 & $\mbox{611299}$ & 2012-Jan-06 & $0.58 $\\
7 & $\mbox{611311}$ & 2012-Jan-18 & $0.62 $\\
8 & $\mbox{611315}$ & 2012-Jan-20 & $0.88 $\\
9 & $\mbox{611319}$ & 2012-Jan-22 & $0.81 $\\
10 & $\mbox{611323}$ & 2012-Jan-24 & $0.67 $\\
11 & $\mbox{611327}$ & 2012-Jan-27 & $0.98 $\\
12 & $\mbox{611331}$ & 2012-Jan-29 & $0.86 $\\
13 & $\mbox{611335}$ & 2012-Feb-02 & $0.86 $\\
14 & $\mbox{611351}$ & 2012-Feb-16 & $0.50 $\\
15 & $\mbox{611355}$ & 2012-Feb-19 & $0.99 $\\
16 & $\mbox{611359}$ & 2012-Feb-21 & $0.79 $\\
17 & $\mbox{611363}$ & 2012-Feb-23 & $0.73 $\\
18 & $\mbox{611367}$ & 2012-Feb-26 & $0.83 $\\
19 & $\mbox{611371}$ & 2012-Feb-29 & $0.90 $\\
20 & $\mbox{611375}$ & 2012-Mar-03 & $0.97 $\\
excluded & $\mbox{611379}$ & 2012-Mar-06 & $0.82 $\\
21 & $\mbox{611387}$ & 2012-Mar-13 & $0.70 $\\
22 & $\mbox{611391}$ & 2012-Mar-15 & $1.08 $\\
23 & $\mbox{611395}$ & 2012-Mar-17 & $0.91 $\\
24 & $\mbox{768813}$ & 2012-May-08 & $0.74 $\\
25 & $\mbox{768817}$ & 2012-May-11 & $0.85 $\\
26 & $\mbox{768820}$ & 2012-May-17 & $0.77 $\\
27 & $\mbox{768826}$ & 2012-May-24 & $1.27 $\\
stacked &  &  & $0.67 $\\
\hline
\end{tabular}
\end{table}

The individual exposures were reduced and combined in single epoch images by making use of the \emph{VST-Tube} pipeline \citep{Grado}, developed for the data produced by the \emph{VST} telescope. A short description of the processing steps follows; details are given in Grado et al. (in preparation).
The instrumental signature removal includes overscan, bias and flat-field correction, CCD gain harmonization, and illumination correction. Hereafter, the procedure secures relative and absolute astrometry, photometric calibrations, and the final co-addition of the images for each epoch. 
The overscan correction is applied by subtracting the median bias value measured over a proper overscan region from the images; the
master-bias is calculated as a sigma-clipped average of bias frames and subtracted from each image.
The master-flat is a combination of a master twilight flat, to correct for pixel-to-pixel sensitivity variation, and a super sky-flat, made from a combination of science images, accounting for low spatial frequency gain variation. The master twilight and master sky-flat are produced using a robust sigma clipped average of overscan and bias corrected twilight frames and science frames, respectively. Wherever possible the master frames have been produced independently for each epoch.

Differences in electronic amplifiers are reflected in different CCD gains. To have the same zero-point for all the mosaic chips, a gain harmonization procedure was applied. The procedure finds the relative CCD gain coefficients which minimize the background level differences in adjacent CCDs. 
The next correction applied to the images is due to scattered light. This is a common feature of wide-field imagers where telescope and instrument baffling is demanding. The scattered light adds a spatially varying component to the background. After flat-fielding, the image background will appear flat, but the photometric response will be position dependent \citep{IC}. Such an error in the flat-fielding can be mitigated through the determination and application of the illumination correction (IC) map. The IC map is obtained by choosing some standard fields and comparing the instrumental magnitudes with those of a reference catalog of stars uniformly spread across the field. Specifically we used the point spread function (PSF) magnitudes from the Data Release 8 \citep{Aihara} of the Sloan Digital Sky Survey (SDSS). The differences in magnitude $\Delta$mag(x,y) as a function of the position are fit with a generalized additive model (GAM; \citealt{gam}) in order to obtain the illumination correction map. The GAM allows us to generate a well-behaved fitting function even when the $\Delta$mag(x,y) does not uniformly sample the whole mosaic and, generally, the resulting fit surface shows smoother behavior at the frame edges compared to a simple surface polynomial fit. After the IC corrected flat-fielding, the images will recover a uniform zero-point over the field, but still the background will not be flat. To achieve a flat background, the IC surface, properly rescaled, is also subtracted from the images. 

The absolute photometric calibration was computed on the photometric night 2011-Dec-17 by comparing the observed magnitudes of stars in photometric standard fields with SDSS photometry. The simultaneous fit for the zero-point and color term gives values of $zp(r)=24.631 \pm 0.006$ and $ct(g-r)=0.040 \pm 0.016$, respectively. The extinction coefficient was taken from the extinction curve M.OMEGACAM.2011-12-01T16:15:04.474 provided by ESO. 
The relative photometric correction among the exposures was obtained by minimizing the quadratic sum of differences in magnitude between overlapping detections. The r.m.s. of the magnitude residuals taking only the sources with a high (> 90) signal-to-noise ratio (S/N) into account is 0.039 mag. The absolute astrometric accuracy compared to the reference 2MASS \citep{2mass} catalog is $0.28\arcsec$, while the relative astrometric accuracy, computed as the quadratic sum of the errors along RA and Dec, is $0.06\arcsec$; the tool used for these tasks is \emph{SCAMP} \citep{scamp}.
The tool \emph{SWARP} \citep{swarp} was used for the image resampling in order to apply the astrometric solution and to produce the final combined, single-epoch, and stacked images, obtained by means of a weighted average.

\section{Selection of variable sources}
\label{section:selection}
For the variability selection we followed an approach similar, but not identical (see below), to the one proposed in \citet{Trevese}.

We extracted a catalog of sources by running \emph{SExtractor} \citep{Bertin}, deriving total, isophotal, and aperture magnitudes through a set of fixed apertures for all epochs. The optimal aperture size for AGN identification should include most of the flux from the central source and minimize the contamination from nearby objects or the host galaxy itself. We therefore selected a 2\arcsec diameter aperture, centered on the source centroid, which on average encloses about $70\%$ of the flux from a point-like source in our catalogs, and we performed an aperture correction to take the effects of seeing into account, by means of growth curves of bright stars across the FoV; the stars chosen are isolated, non-saturated, and distant from regions affected by defects (image edges, reflections, bright star halos, etc.). 

We determined, for each epoch, the ratio of the flux from the reference star enclosed in a 2\arcsec diameter aperture and the flux corresponding to an aperture enclosing $90\%$ of the total flux, thus deriving the aperture correction factor for each epoch, that we then applied to our sources.
The correction technique is based on the assumption that the source is point-like; hence, it does not work reliably with very faint AGNs -- where the host galaxy contribution is considerable even in a 2\arcsec diameter aperture -- because it overestimates variability by improperly correcting part of the flux from the galaxy. We find that the average variability of the most extended sources is $\leq0.01$ mag higher than for the most compact sources as an effect of the over-correction; this contributes to raising the variability threshold, when all the sources (variable and non-variable) are taken into account, returning a conservative sample of optically variable objects. The finding that most of our candidates are compact (Section \ref{section:colors}) confirms that we are not severely overestimating the variability of extended sources. We also tested the alternative method used by \citealt{Trevese} (see also references therein) where each source magnitude is normalized to a reference epoch using the average magnitude of non-variable sources. In principle, the two methods can return different results for extended sources; nonetheless, we verified that this is not the case, since both the results are consistent within the adopted magnitude limit. Our approach is straightforward and easier to apply when dealing with wide-field images, which are characterized by PSF variations across the FoV; on the other hand, the method used in \citet{Trevese} is more sensitive to LLAGNs, but different corrective factors are required for extended sources with different profiles (e.g., early- \emph{vs} late-type galaxies). Thus, although the present paper mostly concentrates on star-like sources, the method is suitable for application to LLAGNs, which constitute a poorly known fraction of the AGN population; a more refined analysis that focuses specifically on the search for LLAGNs may be implemented in a future work.

Since our observations date back to the first period of activity of the \emph{VST}, the data were affected by several aesthetic and electronic problems (as detailed below) due to the lack of knowledge of the corrective factors detailed in the previous section, and of the detector response, most of which were fixed in the following months. For this reason, in this work we decided to use a conservative approach aimed at minimizing the number of spurious sources in the final sample of AGN candidates; hence, we excluded from our analysis all those regions such as the edges (an area approximately ranging from $1^\circ\times125\arcsec$ to $1^\circ\times270\arcsec$ depending on the side) of each epoch, where the S/N is typically very low, together with all the areas affected by the presence of residual satellite tracks or bright star halos, as well as regions affected by scattered light from bright stars just outside the FoV. Furthermore, the weight maps of six epochs revealed that one of the CCDs is affected by sudden and unpredictable variations in its gain factor due to electronic problems\footnote{Recently this flaw has been fixed.}. As a consequence, we excluded the corresponding area from all epochs, to limit the presence of spurious variable sources. The exclusion of additional regions was necessary in three epochs, because of the presence of arcs due to scattered light of bright stars outside the FoV, and, in two cases, because of the contamination by the laser guide star for atmospheric distortion corrections from the nearby \emph{VLT}, which happened to fall in the \emph{VST} FoV\footnote{In the early operation phase the software system designed to prevent this type of problem was not operational yet.}. Overall, we excluded about $23\%$ of the surveyed area. In particular, to remove the stellar halos and diffraction spikes of bright stars we used the masks created by means of the procedure by \citet{Huang}, which automatically accounts for the position of the star inside the FoV and the field orientation. 

The catalogs of sources from each epoch were matched using a matching radius of $1\arcsec$. To ensure a robust light curve variance measurement and to exclude fast transients, we restricted our analysis to the objects detected in at least six epochs (37699 sources). The six epoch threshold is an arbitrary choice, approximately corresponding to a detection in $20\%$ of the epochs (6 out of 27); if modified, the number of sources constituting the sample changes by a few percent, and the corresponding subsample of variable objects changes by just a few units. 

From the light curve of each source \emph{i} we defined an average magnitude $\overline{\mbox{mag}}_i$ and the corresponding r.m.s. deviation \emph{$\sigma_i$},
\begin{equation} 
\overline{\mbox{mag}}_i^{ltc} = \frac{1}{N_{epo}}\sum_{j=1}^{N_{epo}}\mbox{mag}_i^j\mbox{   ,}\qquad\sigma_i^{ltc}={\left[\frac{1}{N_{epo}}\sum_{j=1}^{N_{epo}}{(\mbox{mag}_i^j-\overline{\mbox{mag}}_i^{ltc})}^2\right]}^{\frac{1}{2}}\mbox{,}\label{eqn:avg_stdev}
\end{equation}
$N_{epo}$ being the number of epochs where the source is detected. 
Since the completeness limit of the single epoch catalogs is \emph{r}(AB) $\approx 23$ mag, we limited our analysis to the objects with magnitude $<23$ mag.

In order to select optically variable sources, we computed the running average of the r.m.s. deviation \emph{$\langle\sigma_i^{ltc}\rangle$} and its own r.m.s. deviation r.m.s.$_{\langle\sigma_i^{ltc}\rangle}$ over a $0.5$ mag wide bin, then we defined a variability threshold, so that we assumed an object to be variable if
\begin{equation}
\sigma_i^{ltc} \ge \langle\sigma_i^{ltc}\rangle + 3\times\mbox{r.m.s.}_{\langle\sigma_i^{ltc}\rangle}.\label{eqn:var_threshold}
\end{equation}
The variability significance is thus defined as \citep{Bershady}
\begin{equation}
\sigma^*=\frac{\sigma_i^{ltc}-\langle\sigma_i^{ltc}\rangle}{\mbox{r.m.s.}_{\langle\sigma_i^{ltc}\rangle}}\mbox{   ;}\label{eqn:significance}
\end{equation}
i.e., we select as variable sources those having $\sigma^* \geq 3$.
\begin{figure}[tb]
            {\includegraphics[width=\hsize]{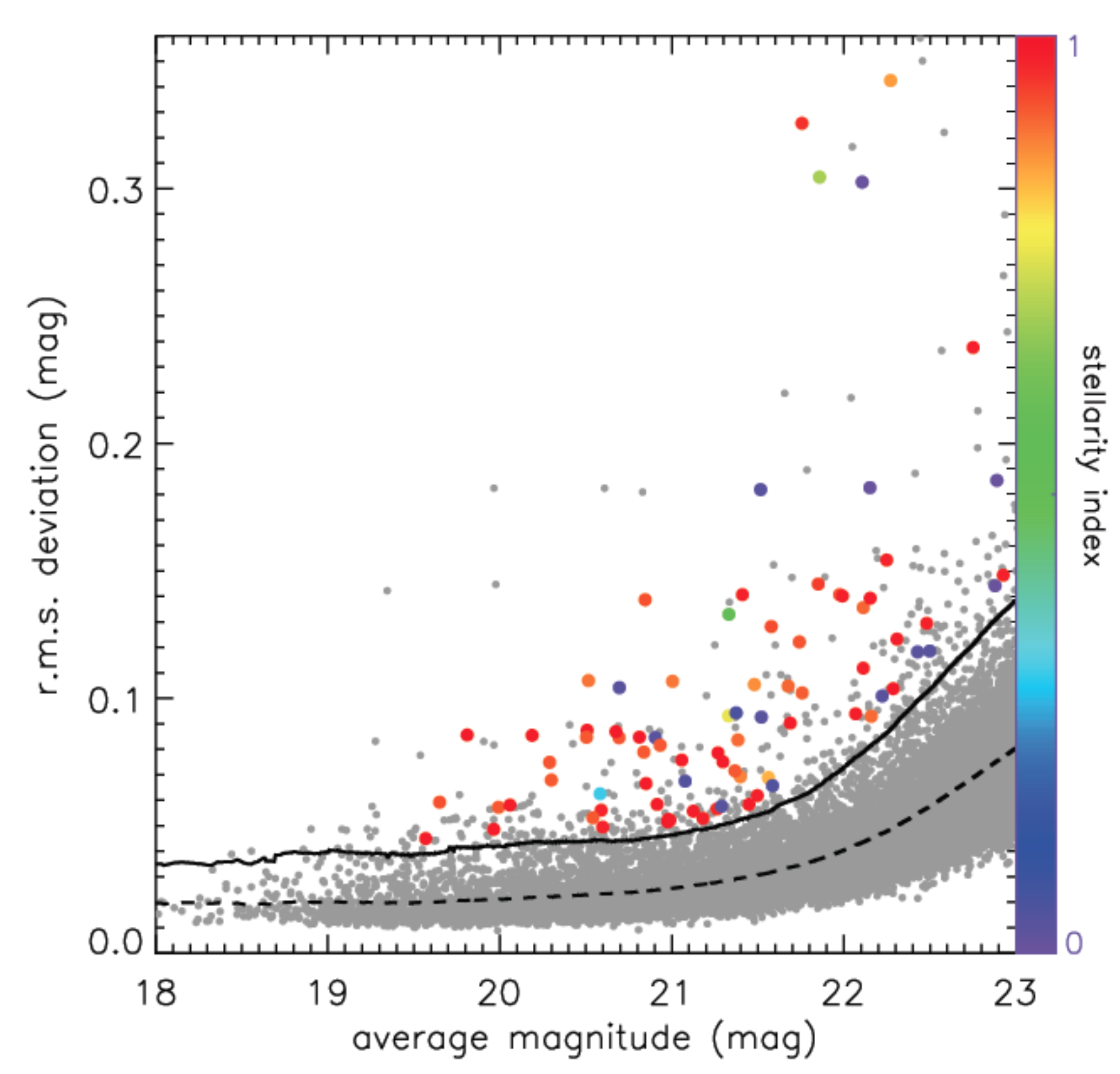}}
\caption{\footnotesize Light curve r.m.s. as a function of magnitude. The running average of the r.m.s. (dashed line) and the variability threshold (solid line) are also shown. Objects above the threshold are assumed to be variable. Large dots represent the sources belonging to our secure sample (end of Section \ref{section:selection}). They are color-coded according to the \emph{HST} stellarity index as in Fig. \ref{fig:rzk}; three sources do not have a \emph{HST} counterpart (see Section \ref{section:discussion}), hence we referred to the \emph{VST} COSMOS stellarity index in those cases.}\label{fig:stdev_vs_mag}
   \end{figure}
The sample of sources with \emph{r}(AB) $< 23$ mag consists of 18282 objects (hereafter \emph{VST} complete sample); 153 ($\approx 1\%$) of them turned out to be optically variable. Figure \ref{fig:stdev_vs_mag} shows the standard deviation \emph{$\sigma_i^{ltc}$} as a function of the average magnitude $\overline{mag}_i^{ltc}$ and the variability threshold (solid line) for the \emph{VST} complete sample. The dashed line represents the running average \emph{$\langle\sigma_i^{ltc}\rangle$} of the r.m.s. deviation.

\begin{figure*}[tb]
 \centering
 \subfigure[quality label 1]
   {\includegraphics[width=10cm]{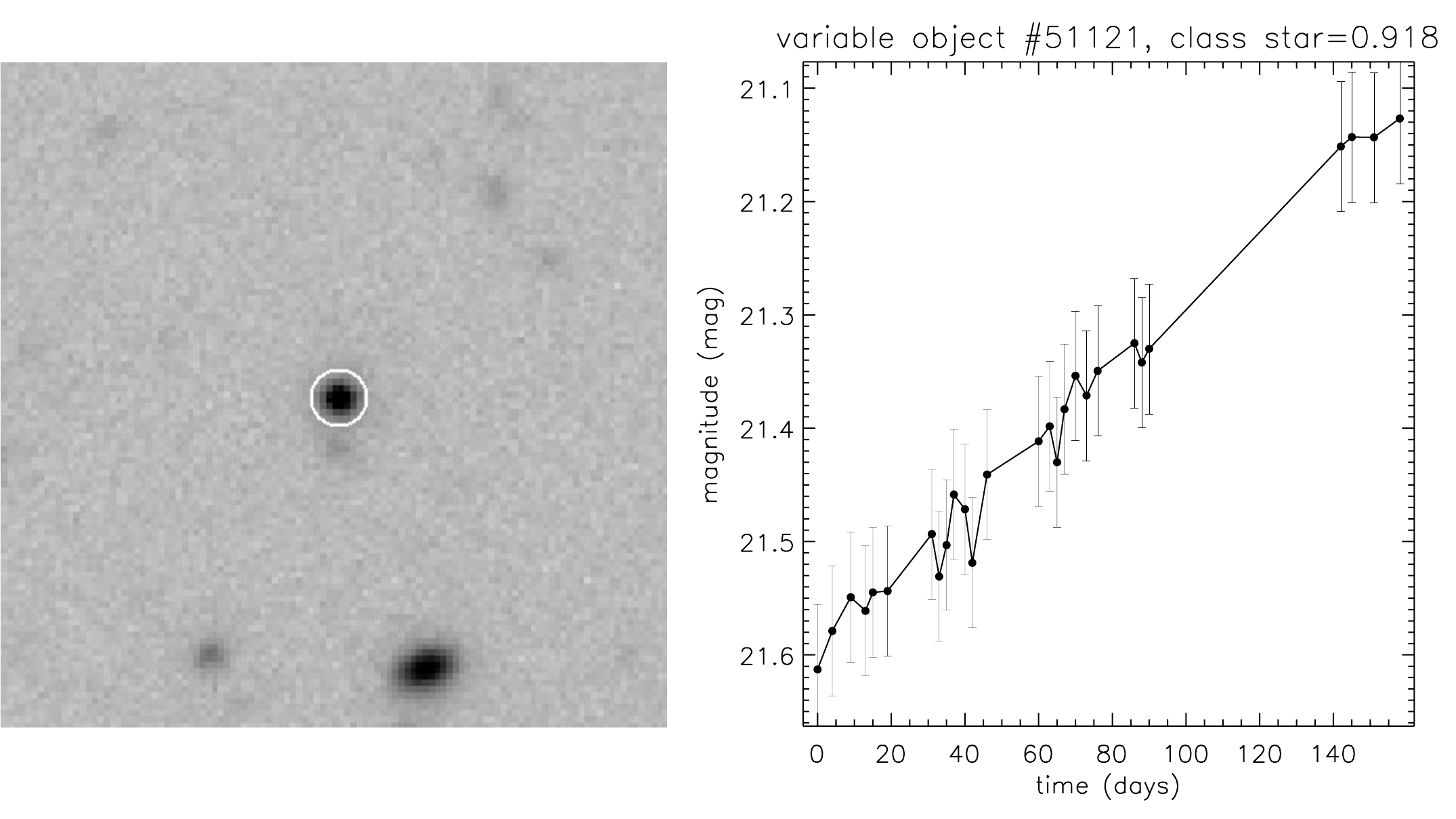}}
 \subfigure[quality label 2]
    {\includegraphics[width=10cm]{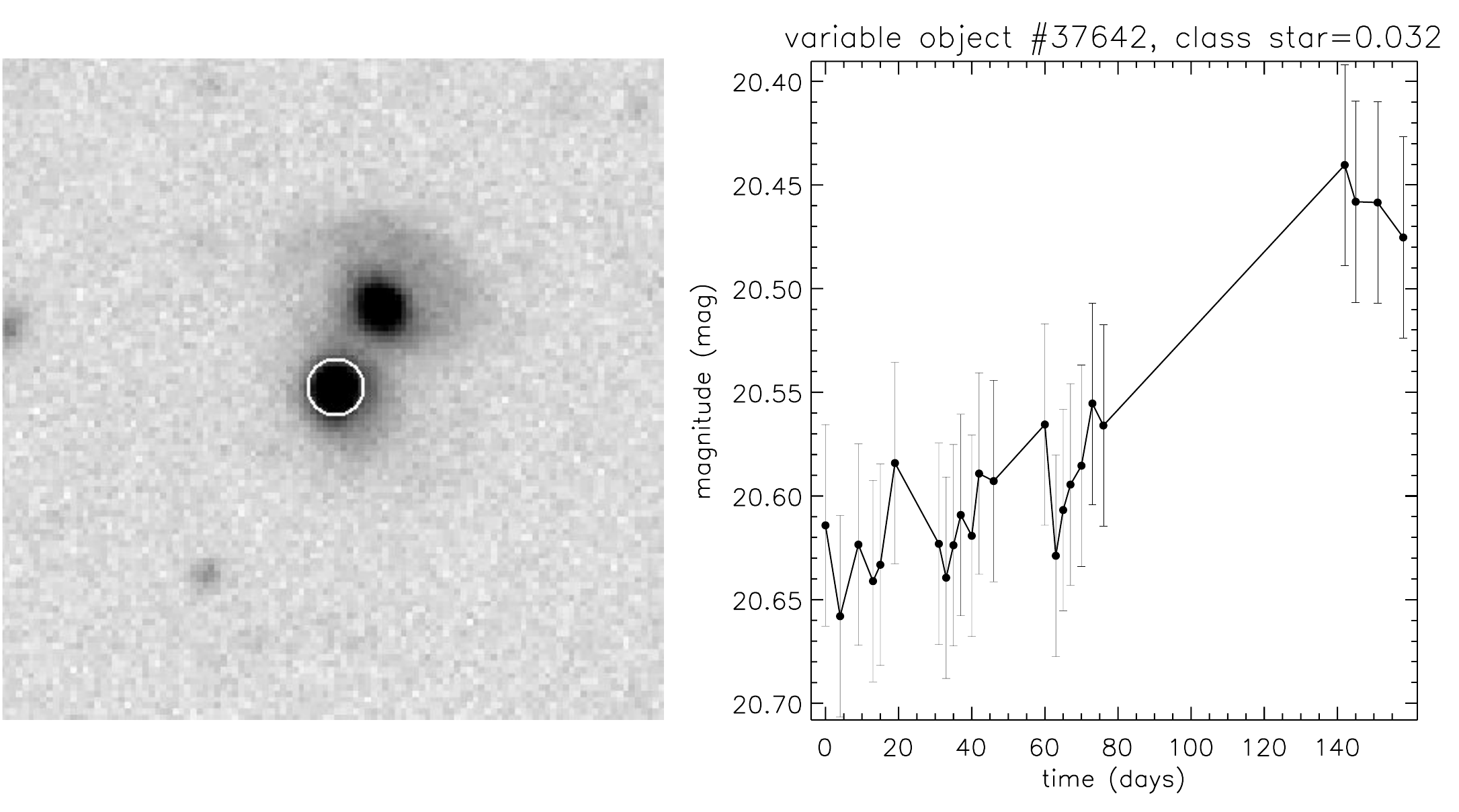}}
   \subfigure[quality label 3]
    {\includegraphics[width=10cm]{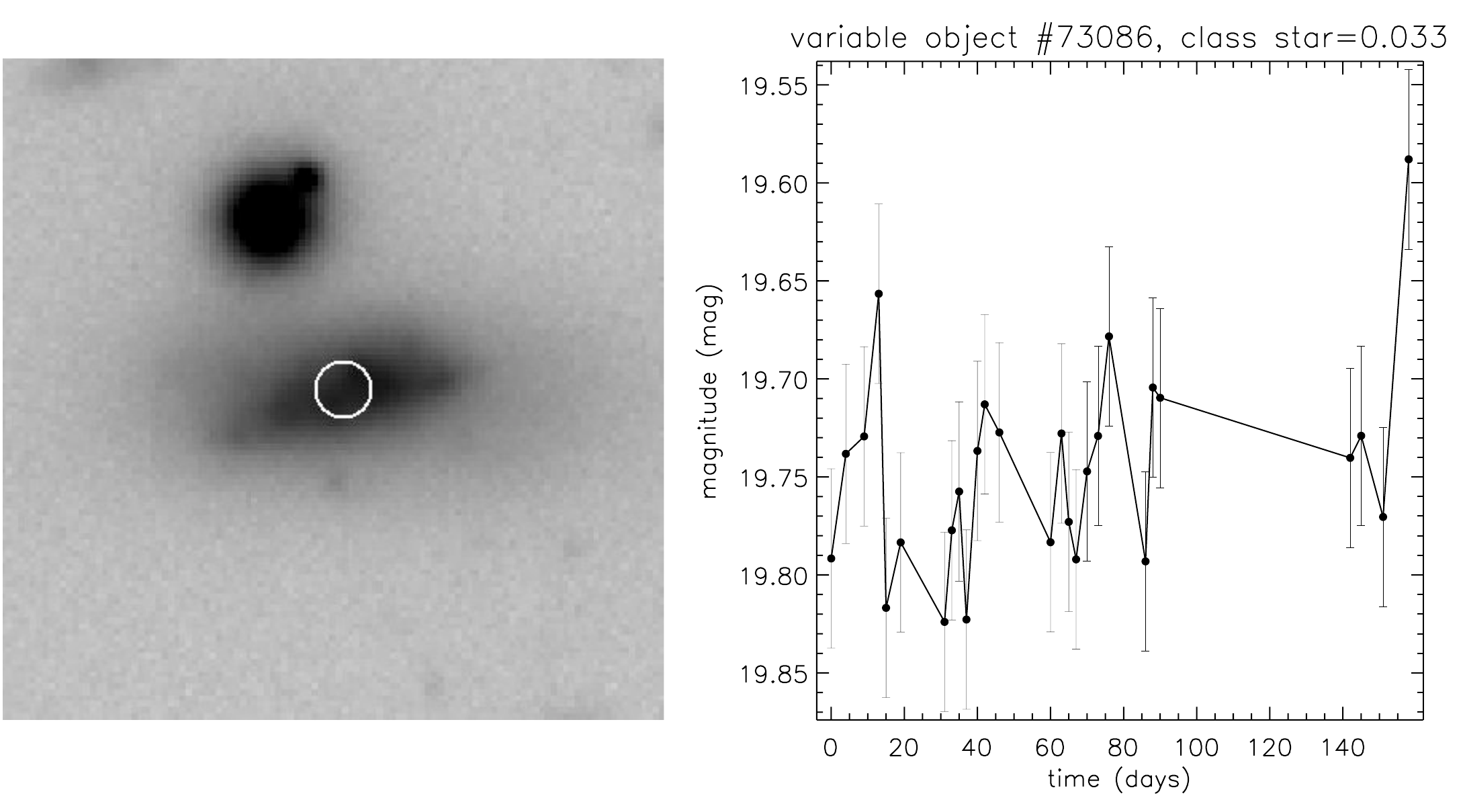}}
   \caption{\footnotesize{Examples (from the stacked image) of variable candidates assigned to different quality classes, with the corresponding light curve. The white circles correspond to the $2\arcsec$ diameter aperture and are centered on the object coordinates listed in the catalog. Objects labeled 1 (panel (a)) are generally isolated and free from aesthetic defects. In the case of the objects belonging to class 2 (panel (b)), potential problems (e.g., the presence of a neighbor) must be taken into account. Objects labeled 3 are probably spurious variable sources: in panel (c) an extended, elongated object with a very bright and close companion is shown as an example. The error bars correspond to our adopted threshold of $\langle\sigma_i^{ltc}\rangle + 3\times\mbox{r.m.s.}_{\langle\sigma_i^{ltc}\rangle}$. The objects in panels (a) and (b) are, respectively, nos. 41 and 71 in Table \ref{tab:secure_sample} (see Section \ref{section:discussion}).}}\label{fig:flag_examples}
\end{figure*}

The sample of 153 optically variable sources includes some objects whose variability is doubtful: sources falling in regions affected by residual aesthetic defects which were not properly masked (hot pixels, stellar diffraction spikes, etc.); very extended objects which may be affected by problems in the centroid identification (e.g., late-type galaxies with irregular morphology) and whose light curve may be affected by the overcorrection problem we mentioned above; objects with a very near and bright companion so that, whether they are deblended or not, it is not possible to establish whether the variability is an intrinsic property of the source or if it is due to PSF variations as a consequence of different seeing conditions, combined with the contamination from the nearby source.
The vast majority of the spurious sources are of the last type. In order to identify and reject the spurious candidates, we visually inspected both the objects and their light curves, attributing to each candidate a quality label ranging from 1 to 3, according to the following criteria:
\begin{enumerate}
\item (70 sources) strong candidate, no evidence of problems or defects;
\item (13 objects) likely variable candidate, potentially affected by the presence of a neighbor, or by minor aesthetic problems;
\item (68 sources) very uncertain variability, likely spurious.
\end{enumerate}
In Fig. \ref{fig:flag_examples} we show an object per class, together with the corresponding light curve, as an example. In the case of close neighbors, we rejected objects with a nearby, point-like source within $2\arcsec$ (centroid-to-centroid distance) when their magnitude was $\leq \mbox{mag}_{source}+1.5$; we also excluded five sources which happened to fall in the halo of extended, saturated objects. We point out that our choice to visually inspect all candidates is due to the need to understand the variety of problems that can affect variability measurements in the \emph{VST} wide-field images; for the future, the rejection criteria can be partly automated (as we already did with star halo masking) for future large scale surveys (e.g., \emph{LSST}). In addition, different variability measurement approaches, such as the PSF-matched image subtraction method (discussed in Section \ref{section:SN}) may also overcome some of these issues, as the close neighbor contamination, at the expense of reducing the S/N of the central AGN (due to PSF degradation).
The variability analysis that we describe in the next section is limited to the sources labeled 1 or 2 (hereafter secure sample), made up of 83 sources and hence constituting $54\%$ of the initial sample of 153 AGN candidates with mag $< 23$; we will mention Type 3 sources only when appropriate.

\section{The nature of variable sources}
\label{section:validation}
In the present section we investigate the nature and properties of our secure sample in order to distinguish different classes of objects and study their features. The validation of our candidates follows, in some cases, from an already available classification published in other catalogs of COSMOS sources; furthermore, when no prior classification is available, we rely on properties derived from \emph{VST} data themselves (r.m.s. variability, light curve, optical morphology), coupled with additional diagnostics derived from the multiwavelength database.

\subsection{Supernova identification}
\label{section:SN}
\begin{figure*}[htb]
 \centering
 \subfigure[C-COSMOS Identification catalog \citep{Civano} and \emph{VST}-COSMOS field.]
   {\includegraphics[width=8cm]{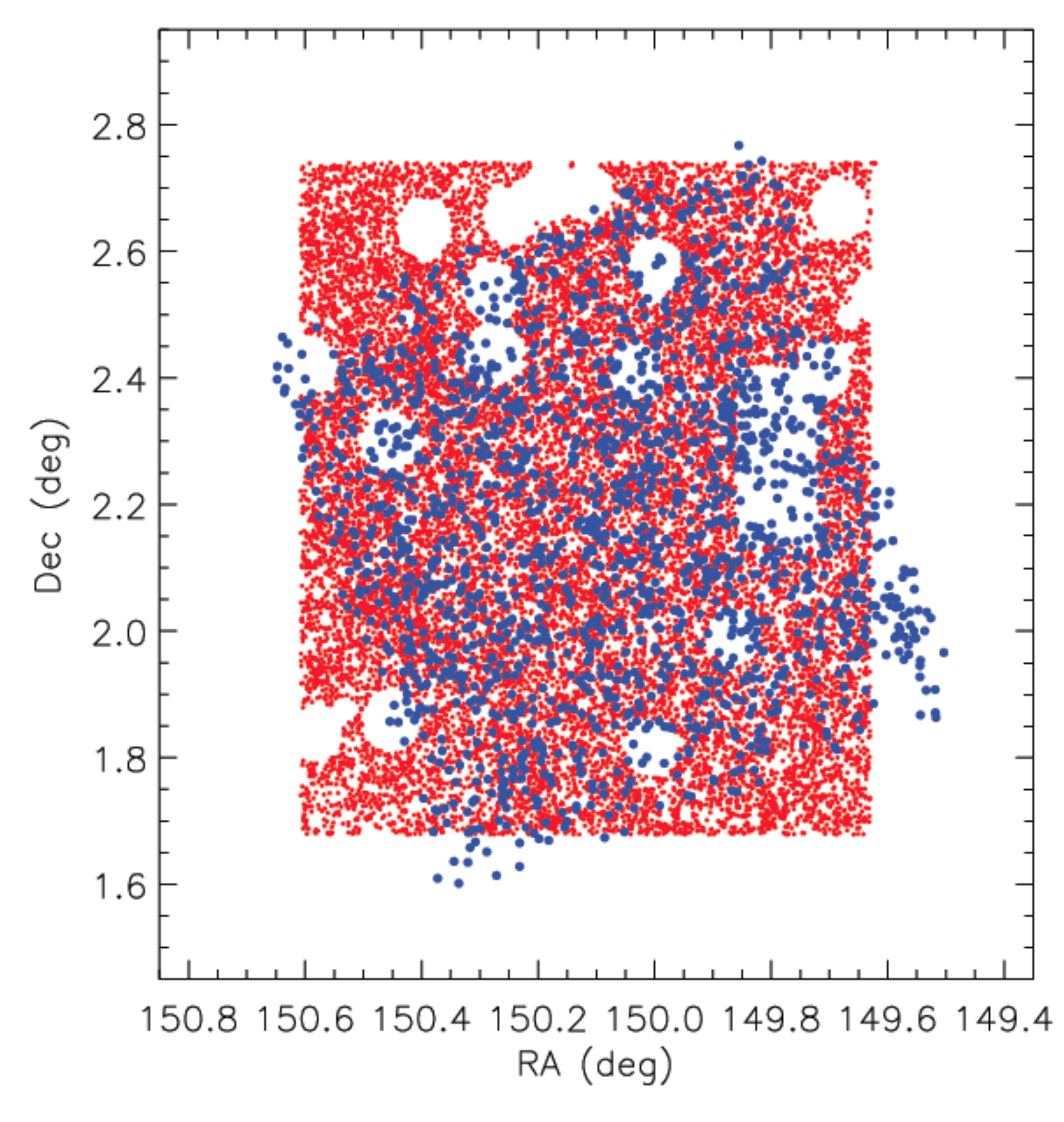}}
 \hspace{1cm}
 \subfigure[\emph{XMM}-COSMOS Point-like Source catalog \citep{Brusa} and \emph{VST}-COSMOS field.]
    {\includegraphics[width=8cm]{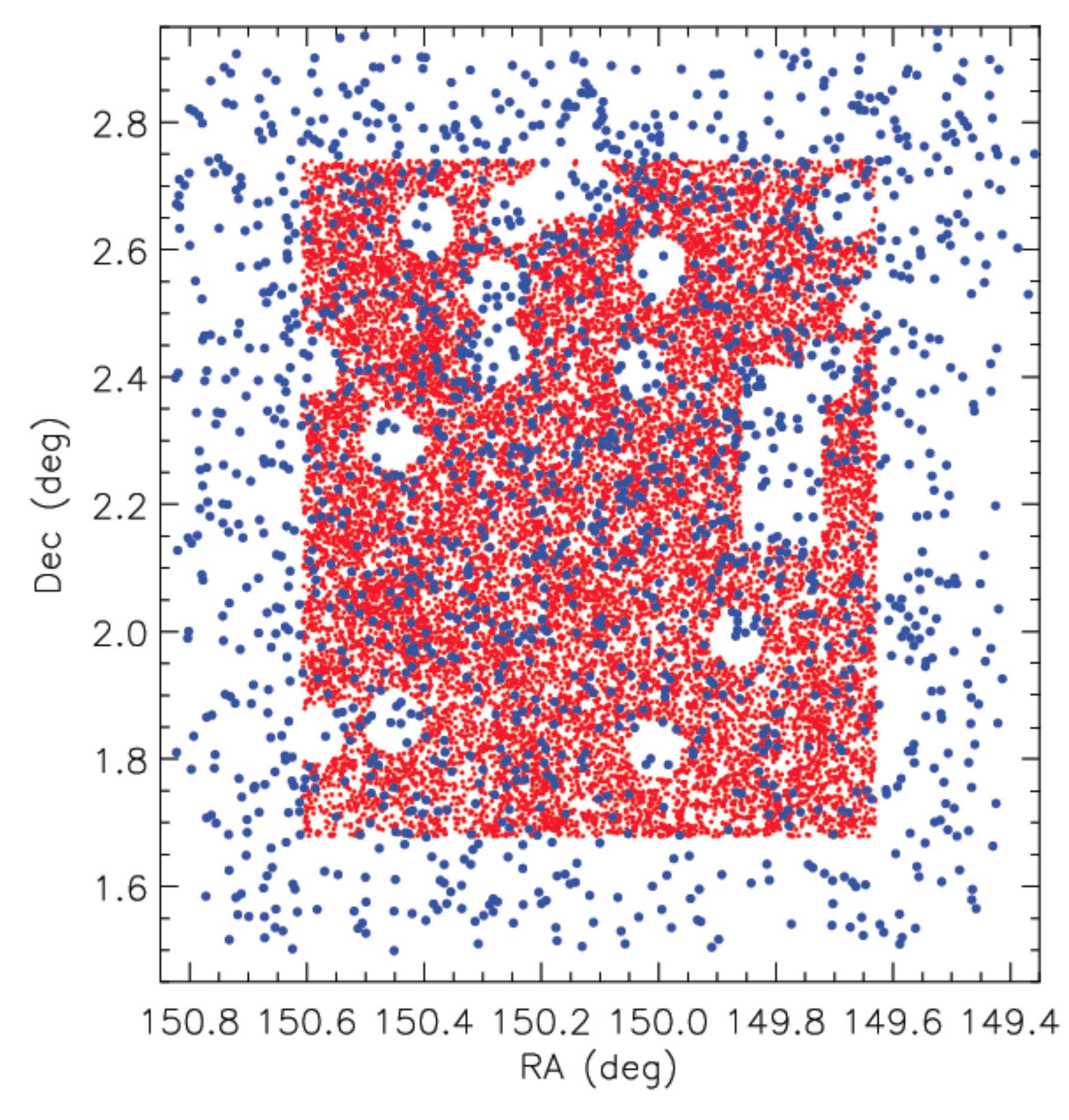}}
   \caption{\footnotesize{Comparison between the \emph{VST} COSMOS catalog (smaller red dots) and the \emph{Chandra} (panel (a)) and \emph{XMM} (panel (b)) COSMOS catalogs (larger blue dots). The holes in the \emph{VST} catalog represent regions that were masked and hence not taken into account in our variability analysis because of the presence of bright star halos (see Section \ref{section:selection}). It is apparent that part of the COSMOS field as imaged by the \emph{VST} is not covered by \emph{Chandra} observations. In both panels we showed all the sources in each of the X-ray catalogs, but we limited our analysis to those falling in the \emph{VST} FoV, out of the masked areas.}}\label{fig:fields}
\end{figure*}
We already mentioned that the dataset used for our analysis comes from a SN survey program, whose primary science goal was to measure the rate of the different types of SNe at medium ($0.3-0.8$) redshifts. This dictated the observing strategy, namely the choice of filters, exposure time, and cadence. While the SN search project is still in progress (observations will be completed by mid 2015), the analysis of the data obtained so far has been completed and will be described in detail in a dedicated paper \citep{Cappellaro}. In order to search for SN candidates, the calibrated mosaic images produced by \emph{VST-Tube} were processed through a dedicated pipeline. Here we briefly outline the dedicated SN search pipeline whose results we cross-correlated to our variable source sample. The SN candidates were identified in the difference images obtained by subtracting, from the image taken at a given epoch, a reference image taken a few months earlier. Transients are positive sources in the difference, but, in practice, most of the detections are artifacts (defects on one of the two images, poorly subtracted residuals of bright stars, small scale astrometric mismatches, etc.). Most of the false transients are rejected by requiring specific constraints on the source metrics (FWHM, flux radius, location compared to bright source, etc.), but even in this case the final selection is performed by visual inspection. The selected transients are classified on the basis of the observed light curves. The frequent monitoring allows derivation of well-sampled light curves and, although less frequent, color measurements are also very useful. In general, SNe and variable AGNs are easily separated because the latter have erratic light curves on long timescales. However, in a few cases the distinction between AGNs and SNe, especially for Type IIn SNe with slowly evolving light curves, may be impossible especially if detection occurs at the edge of the search window, i.e., in the very early or late epochs. Supernova classification is obtained by comparing the observed light curves in different bands to those of template objects, allowing for three free parameters, which are redshift, epoch of explosion, and extinction. The fit is facilitated by constraining the redshift to the range of uncertainty of the host galaxy redshift that for the COSMOS field is available to very deep limits through photometric techniques \citep{Muzzin}. All together, in the observing season 28 SNe were classified (with an additional nine classified as uncertain) by fitting the light curves with different templates. At the same time, about 80 events were labeled as variable AGNs mainly on the basis of their erratic light curves and association with a QSO or galaxy nucleus. It is worth noting that the transient selection algorithm of this search is optimized for SNe and therefore is not expected to be complete or robust for the identification of variable AGNs. However, we point out that all the events that are labeled as possible AGNs by Cappellaro and collaborators and that satisfy our selection criteria (detection in 6+ epochs, magnitude \emph{r}(AB) $< 23$ mag, location in non-masked areas) are classified as AGNs in our work; on the other hand, of our list of 83 optically variable sources, $87\%$ are also found in the list of transients by Cappellaro and collaborators.

We took advantage of the results from the SN pipeline to identify the SNe in our secure sample. A visual inspection of the light curves of the 83 sources in our sample showed ten objects with typical SN light curves, hence we marked them as possible SNe; we then cross-matched our list of possible SNe to the sample of classified sources from the SN search by Cappellaro and collaborators, and found that eight of them were classified as SNe as well, while the remaining sources are not in their list. There are two additional objects belonging to our secure sample and classified as a SN and a possible SN by Cappellaro and collaborators; the light curves that we obtained for those two sources alone do not allow any guesses about their nature. Hereafter, we will label as SNe all eight sources identified in both works, plus the one classified as SN by Cappellaro and collaborators, while the other three (two from our classification plus one from their list) will be considered as possible SNe. A detailed list of the sources in our secure sample, including the SNe, can be found in Table \ref{tab:secure_sample} at the end of this paper.

\subsection{X-ray counterparts}
\label{section:X-rays}
\begin{figure*}[tb]
 \center
 \subfigure[Soft ($0.5-2$ keV) X-ray luminosity \emph{vs} redshift $z$ for our variable sources, compared to the overall X-ray population.]
   {\includegraphics[width=8.5cm]{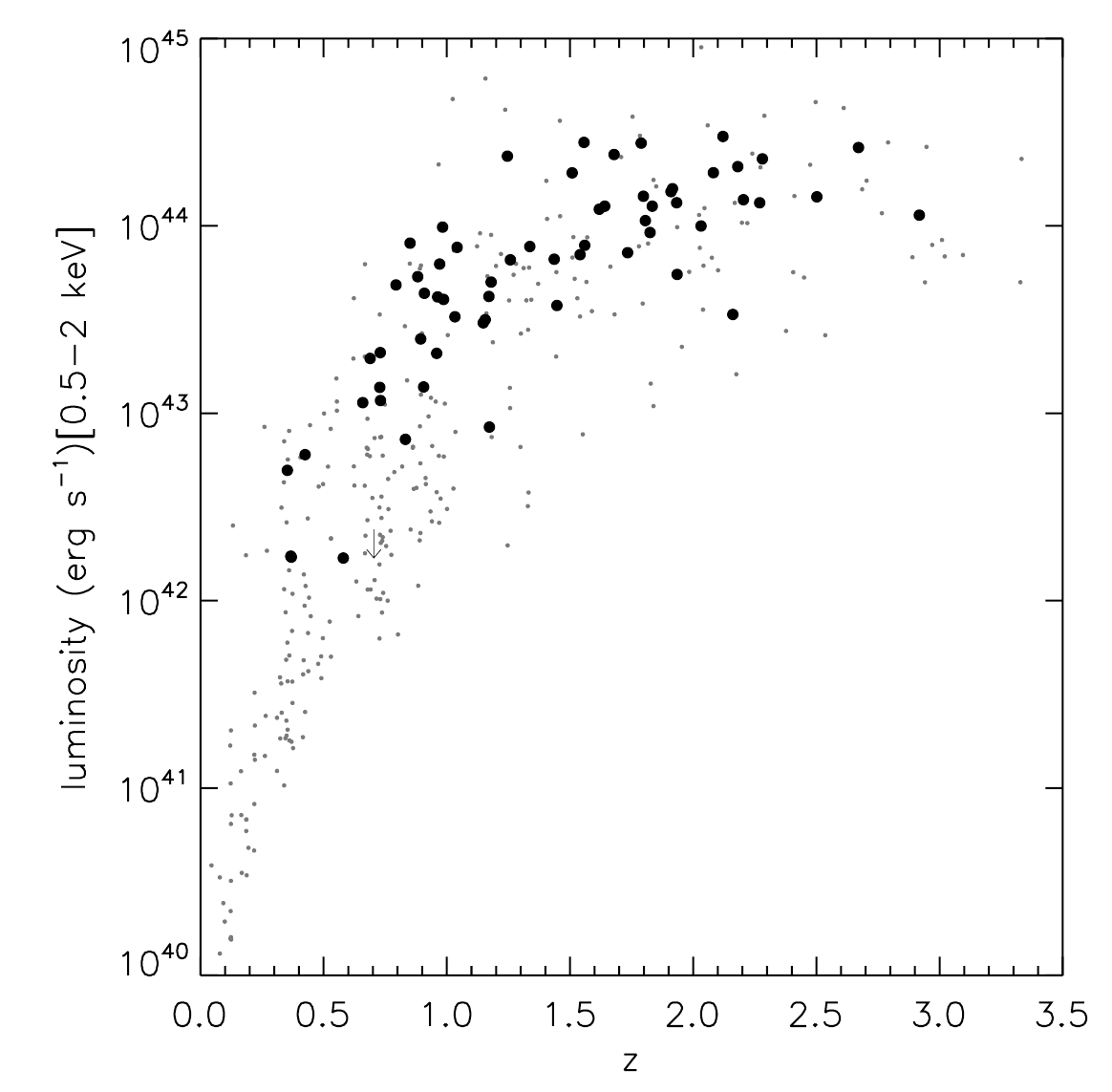}}
 \hspace{1cm}
 \subfigure[Hard ($2-10$ keV) X-ray luminosity \emph{vs} redshift $z$ for our variable sources, compared to the overall X-ray population.]
    {\includegraphics[width=8.5cm]{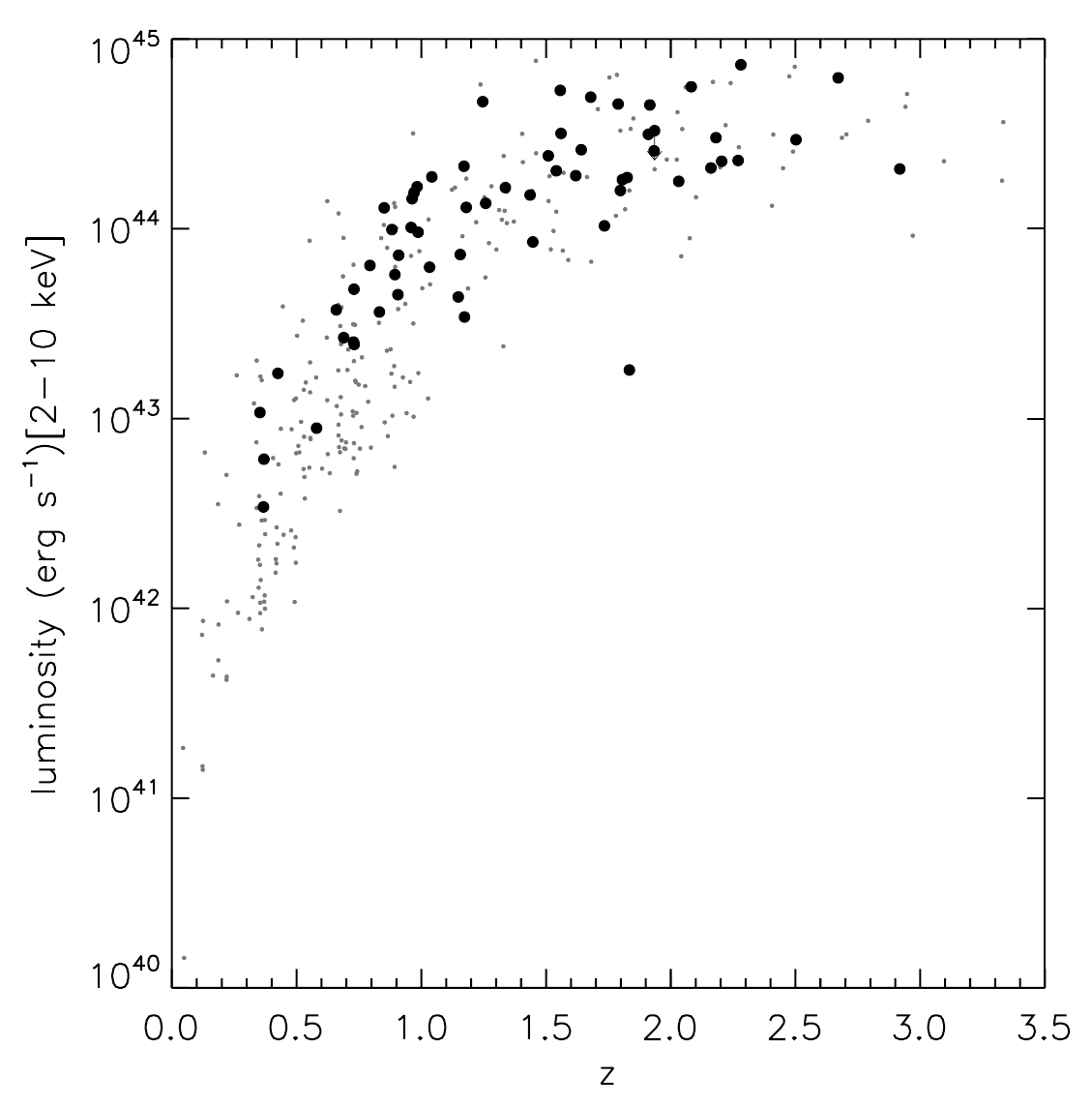}}
\caption{\footnotesize Larger dots represent the 63 optically variable sources with an X-ray counterpart. \emph{Chandra} luminosities and redshifts were used when available, and \emph{XMM} data were adopted for the remaining sources. A spectroscopic redshift value was available for all but three sources, for which photometric redshifts were used \citep{Salvato1}. The downward arrows stand for those sources for which only upper limits of the flux values were available. Smaller grey dots are from the \emph{Chandra} catalog and represent a reference population.}\label{fig:Lx_vs_z}
\end{figure*}
\begin{figure*}[tb]
 \center
 \subfigure[Soft ($0.5-2$ keV) X-ray to optical flux ratio.]
   {\includegraphics[width=8.5cm]{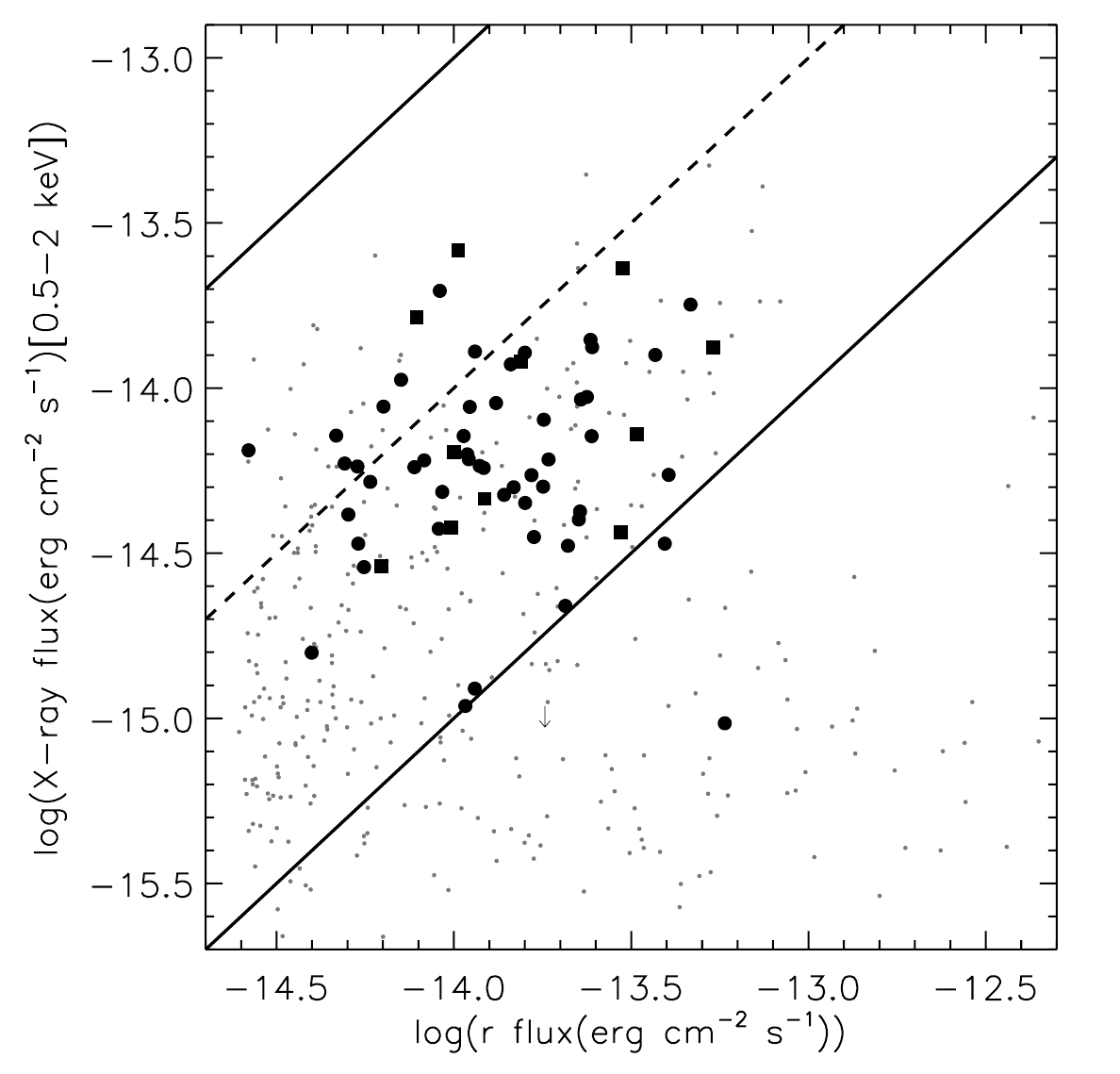}}
 \hspace{1cm}
 \subfigure[Hard ($2-10$ keV) X-ray to optical flux ratio.]
    {\includegraphics[width=8.5cm]{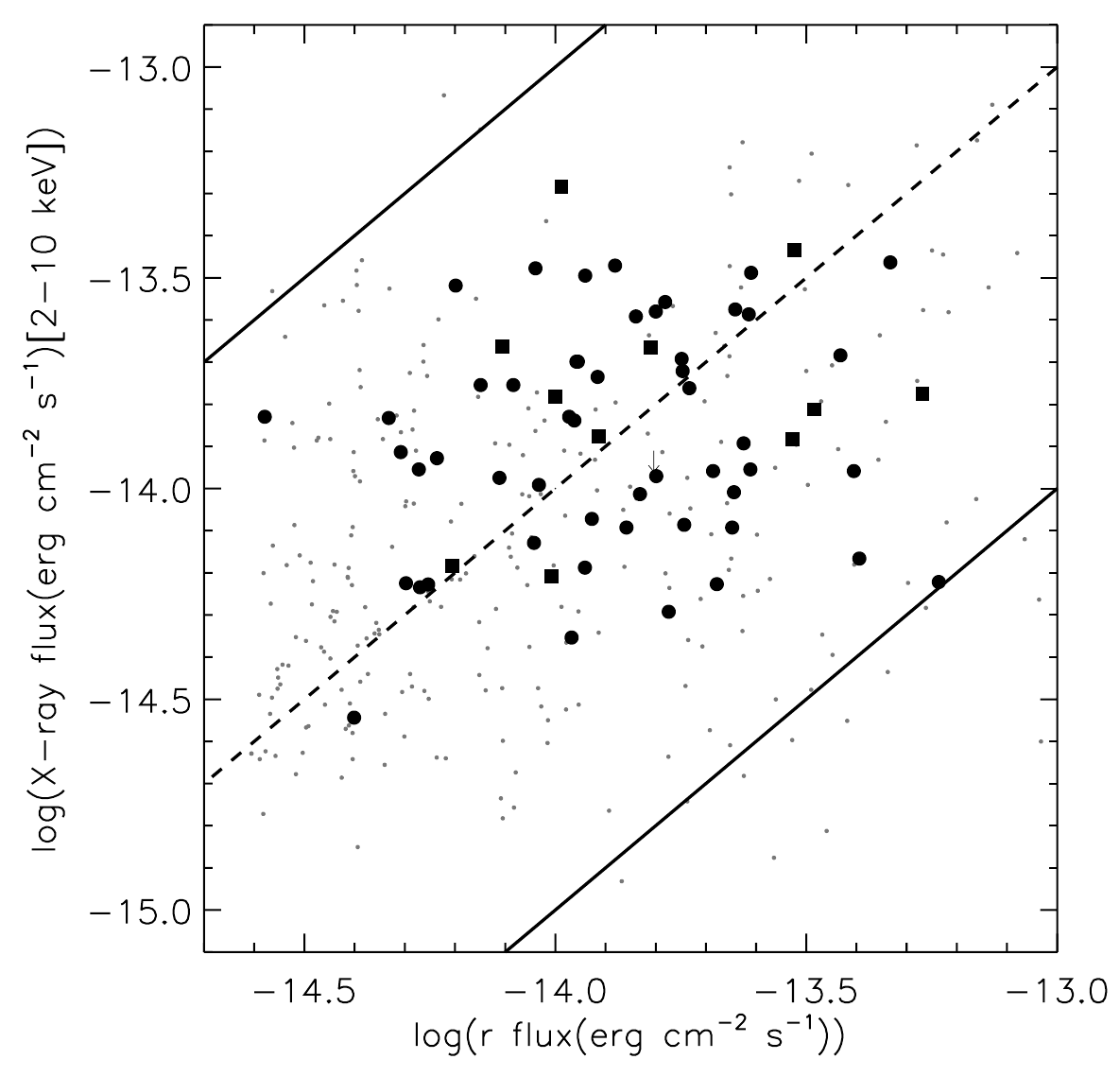}}
\caption{\footnotesize X-ray \emph{vs} optical flux for the \emph{VST} variable objects with an X-ray counterpart from \emph{Chandra} (dots) or \emph{XMM} (squares) catalogs; \emph{Chandra} fluxes were used when available, while \emph{XMM} fluxes were adopted for the remaining objects. Downward arrows stand for those sources for which only upper limits of the X-ray flux were available. The smaller grey symbols in the background are shown as a reference population and stand for the X-ray objects in the \emph{Chandra} catalog having a \emph{VST} counterpart and \emph{r}(AB) $< 23$ mag. The dashed line corresponds to X/O $=0$, the lower and upper solid lines represent X/O $=-1\mbox{ and X/O }=1$, respectively. AGNs typically place themselves in the range where $-1 \le$ X/O $\le 1$.}\label{fig:XO}
\end{figure*}

The presence of X-ray emission, especially when coupled with variability, constitutes strong evidence of the presence of AGN activity. This is why, when investigating the properties of our sample, we made wide use of the available X-ray catalogs of COSMOS objects:
\begin{itemize}
\item the \emph{Chandra}-COSMOS Identification catalog \citep{Civano}, containing 1761 X-ray sources spread over a 0.9 square degree area, with a 160 ks depth in the inner region (0.5 square degrees), and 80 ks depth in the remaining 0.4 square degrees \citep{Elvis}. The limiting depths in soft ($0.5-2$ keV) and hard ($2-10$ keV) X-rays correspond to fluxes of $1.9\times10^{-16}$ erg cm$^{-2}$ s$^{-1}$ and $7.3\times10^{-16}$ erg cm$^{-2}$ s$^{-1}$, respectively, while the depth for the full ($0.5-10$ keV) X-ray band is $5.7\times10^{-16}$ erg cm$^{-2}$ s$^{-1}$. The catalog provides, among other data, optical counterparts for the 1761 sources; it also includes a spectroscopic classification (BLAGN, non-BLAGN, star) for about half the sources in the catalog, and a photometric classification through SED fitting for $94\%$ of the objects;
\item the \emph{XMM}-COSMOS Point-like Source catalog \citep{Brusa}, made up of 1674 X-ray sources. The corresponding program is shallower (60 ks), but wider (2 square degree area) than \emph{Chandra}; the catalog has a flux limit of $\approx1.7\times10^{-15}$ erg cm$^{-2}$ s$^{-1}$, $\approx9.3\times10^{-15}$ erg cm$^{-2}$ s$^{-1}$ and $\approx1.3\times10^{-14}$ erg cm$^{-2}$ s$^{-1}$ over $90\%$ of the area, in the $0.5-2$ keV, $2-10$ keV and $5-10$ keV energy bands, respectively. Spectroscopic classification is provided for approximately half the sample, and a best-fit SED template by \citet{Salvato} was found for $97\%$ of the objects. There is also an additional catalog of 545 Type 1 AGNs \citep{Lusso} from the \emph{XMM}-COSMOS survey, made up of $\sim60\%$ spectroscopically confirmed AGNs; for the others, a reliable photometric estimate of their redshift exists, and their AGN nature is confirmed by their broadband SEDs. 
\end{itemize}
The COSMOS fields as surveyed by different observatories generally do not overlap perfectly. In Fig. \ref{fig:fields} we show the superposition of the fields as imaged by the \emph{VST} and by both the \emph{Chandra} (panel (a)) and \emph{XMM} (panel (b)) telescopes.

On the whole, the X-ray catalogs provide information about 2628 X-ray emitters, of which 1517 fall in the \emph{VST} FoV, in non-masked areas. To investigate the nature of our sample of variable sources, we matched our \emph{VST} complete sample to the optical counterparts of the X-ray sources (as derived in \citealt{Capak} and \citealt{Ilbert1}; see also \citealt{Brusa} and references therein for the counterparts of \emph{XMM} sources) and brighter than \emph{r}(AB) $= 23$ mag; this subsample of X-ray emitters consists of 548 objects (hereafter X-ray sample). 
The rest of the unmatched X-ray sources are missed because:
\begin{itemize}
\item 526 have an optical \emph{VST} counterpart, but the magnitude of their counterpart is \emph{r}(AB) $> 23$ mag, i.e., beyond the completeness limit of our single-epoch catalogs;
\item one source has an optical \emph{VST} counterpart and \emph{r}(AB) $< 23$ mag, but the counterpart is detected in less than six epochs;
\item 442 do not have an optical \emph{VST} counterpart in the 6+ epoch catalog, or in any of the single epoch catalogs. Both the \emph{Chandra} and \emph{XMM} catalogs are at least two magnitudes deeper than ours, hence we certainly miss the fainter sources in the field. Since our analysis is limited to the objects with \emph{r}(AB) $< 23$ mag, we matched the list of 442 objects with several COSMOS optical catalogs (e.g., \citealt{Capak, Ilbert}) providing measures of Subaru \emph{r}(AB) magnitudes and SDSS \emph{r}(AB) magnitudes, and we found that all but 30 out of the 442 sources have optical magnitudes fainter than 25 mag (i.e., the limiting magnitude of our single epoch catalogs). With respect to the 30 objects, we noticed that they are generally on the edge of a masked region or very close to a brighter source, where the completeness is lower, thus they are likely missed because of incompleteness.
\end{itemize}
To summarize, the \emph{VST} sources with \emph{r}(AB) $< 23$ mag and with an X-ray counterpart are 548 out of 18282 ($3\%$); among them, 63 belong to our secure sample, so we can state that $76\%$ of the secure sample is made up of X-ray sources and also that the X-ray sources with an optically variable counterpart are $11\%$ of 548; their X-ray emission, coupled with their optical variability, is a clue to their AGN nature. 
Further evidence comes if we look at their X-ray luminosity, both in the $0.5-2$ and $2-10$ keV bands: it is well known \citep[e.g.,][]{Brandt&Hasinger} that generally non-active galaxies have X-ray luminosities below $10^{42}$ erg s$^{-1}$. Figure \ref{fig:Lx_vs_z} shows the plot of X-ray luminosity $L_{\scriptscriptstyle X}$ as a function of redshift $z$ for the 63 optically variable sources with an X-ray counterpart: for all of them $L_{\scriptscriptstyle X} > 10^{42}$ erg s$^{-1}$, hence we can be confident that they all are AGNs. It is worth noting that only 3 of the 68 sources labeled as 3, and therefore excluded from our analysis, have an X-ray counterpart.

Several types of X-ray sources can be classified on the basis of their X-ray to optical flux ratio, which is defined as \citep{Maccacaro}
\begin{equation}
X/O=\log(f_{\scriptscriptstyle{X}}/f_{opt})=\log f_{\scriptscriptstyle{X}} + \frac{\mbox{mag}_{opt}}{2.5} + C \mbox{ ,}\label{eqn:XO}
\end{equation}
where $f_{\scriptscriptstyle{X}}$ is the X-ray flux measured in the chosen energy range, $\mbox{mag}_{opt}$ is the optical magnitude at the chosen wavelength, and $C$ is a constant which depends on the magnitude system adopted for the observations. Typically, AGNs are characterized by $-1\le$ X/O $\le 1$, while stars and non-active galaxies generally have X/O $< -2$ \citep[e.g.,][]{Mainieri, Xue}.

In Fig. \ref{fig:XO} we show the soft ($0.5-2$ keV; panel (a)) and hard ($2-10$ keV; panel (b)) X-ray flux \emph{vs} \emph{r}-band flux for all the AGN candidates in our list. The large symbols represent the AGN candidates in our sample; the two solid lines define the region where $-1 \le$ X/O $\le 1$. In panel (a) there are three sources lying outside the AGN region; in any case, they place themselves within the AGN locus when we examine hard X-ray \emph{vs} optical fluxes; therefore, coupling the results from the two diagrams, we can state that all 63 sources lie in the region where $-1 \le$ X/O $\le 1$, and so they all are likely AGNs. This means that, if no more information about the nature of the sources were available, we could be confident that $76\%$ (63 out of 83) of the objects in our secure sample are AGNs on the basis of the X/O diagrams and of the X-ray luminosity of the 63 sources. We point out that the X/O of a source is always defined with an uncertainty due to the intrinsic variability of the source combined to the non-simultaneity of the X-ray and optical observations.

\subsection{Spectral properties}
\label{section:sp}
To probe the nature of our variable sources we looked at their spectral properties. We already mentioned (see Section \ref{section:X-rays}) that a spectroscopic and/or photometric classification is available for most of the sources in each of the X-ray catalogs.
The classification in the \emph{Chandra} catalog was derived from spectra, when available, or through template fitting of the broadband SED \citep{Salvato1}, as described in \citet{Civano}. Objects spectroscopically classified are labeled as BLAGNs, non-BLAGNs, and stars; non-BLAGNs could be obscured AGNs as well as non-active galaxies; objects with a photometric classification from SED fitting are divided into unobscured AGNs, obscured AGNs, and galaxies.

In the \emph{XMM} catalog, sources are classified as BLAGNs\footnote{In both catalogs, a source is labeled as BLAGN if its spectrum shows at least one broad (FWHM $> 2000$ km s$^{-1}$) emission line.}, narrow-line AGNs (NLAGNs)\footnote{Sources with unresolved high-ionization emission lines with line ratios suggesting AGN activity, or with rest-frame hard X-ray luminosity $L_{\scriptscriptstyle X}>2\times10^{42}$ erg s$^{-1}$.}, and normal (meaning ``non-active'') galaxies\footnote{Sources with spectra consistent with those of star-forming or normal galaxies, or with rest-frame hard $L_{\scriptscriptstyle X}<2\times10^{42}$ erg s$^{-1}$, or not detected in the hard band.}; part of the best-fit SED templates correspond to Type 1 and Type 2 AGNs.

In total, there are 341 X-ray sources which are confirmed AGNs on the basis of the classifications (spectroscopic or photometric through SED fitting) given in the X-ray catalogs. One classification at least is provided for each of the 63 optically variable objects with an X-ray counterpart. After a merger of the various classifications, assuming that the spectroscopic ones are the most reliable, we can state that the sample of 341 confirmed AGNs is made up of 215 Type 1 AGNs and 126 Type 2 AGNs. 
Specifically, to label the sources in our sample as Type 1 or 2 AGNs we referred to the spectroscopic classification and, when not available, we took into account the photometric classification. The criterion adopted to define Type 1 AGNs is the same in \citet{Brusa} and \citet{Civano} while, in the case of Type 2 AGNs, the label ``non-BLAGN'' from the catalog by \citet{Civano} alone is not sufficient to classify a source. As a consequence, we labeled as Type 2 AGNs all the sources spectroscopically classified as NLAGNs in the catalog by \citet{Brusa}, while we did not classify as Type 2 AGNs any of the non-BLAGNs in \citet{Civano} if no additional classification was available; only one source was labeled as a Type 2 AGN after the photometric classification by \citet{Civano}, as this was the only classification available. In the case of conflicting labels (only two sources) we chose to adopt the classification by \citet{Civano} as it is more recent.

Sixty-two of the X-ray sources that are also optically variable are classified as Type 1 or Type 2 AGNs; in particular, we found 55 ($89\%$ of 62) Type 1 and 7 ($11\%$ of 63) Type 2 AGNs; the remaining X-ray object is classified as a galaxy after its SED, but we note that its X-ray luminosity is $L_{\scriptscriptstyle X}>10^{42}$ erg s$^{-1}$ and that the source lies in the AGN region in both the soft and the hard X-ray \emph{vs} optical flux diagrams, so we can consider this as evidence that the source hosts an AGN. The spectrum of this source is of low quality but does not show evident emission line features. It could thus represent an X-ray bright optically normal galaxy (XBONG; see \citealt{Comastri}). Two XBONGs were also found by \citet{Trevese}, and their nature is still a subject of debate (see \citealt{Malizia}, who mention possible different interpretations).

\subsection{Photometric classification} 
\label{section:colors}
\begin{figure*}[bt]
 \center
            {\includegraphics[width=14cm]{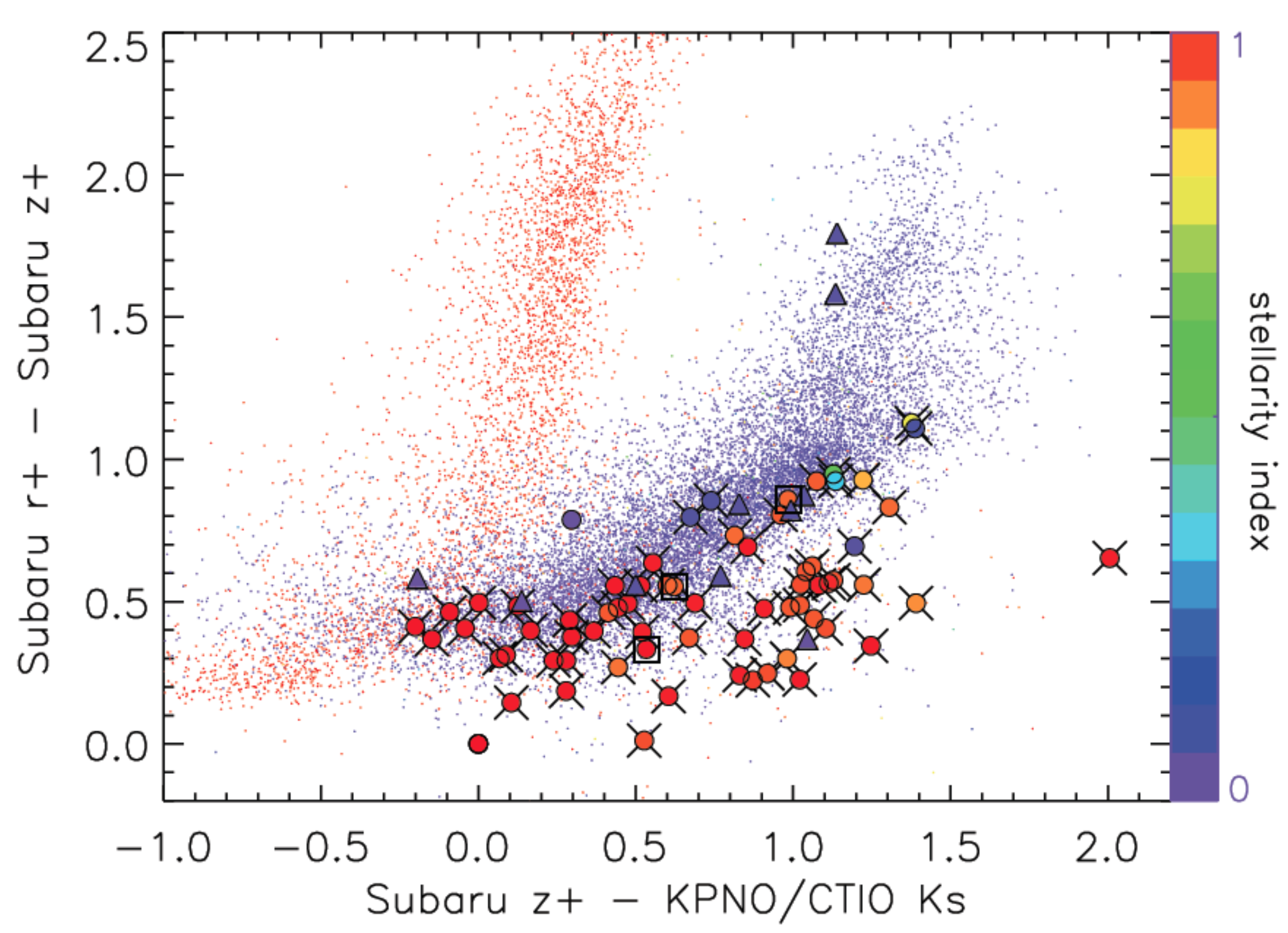}}
\caption{\footnotesize \emph{r-z vs z-k} diagram for 67 out of 71 AGN candidates (larger dots) in our list, for which a stellarity index (ranging from 0 to 1 with decreasing extension of the source: see legend on the right of the plot) and color information are available after other COSMOS catalogs. The faint, smaller objects represent all the objects detected in the \emph{VST} survey of the COSMOS field for which stellarity index and color information are available, and are shown as a reference population. The sources labeled by a cross are the AGNs whose nature has already been confirmed, while the triangles are for the eight SNe common to our catalog and to that by Cappellaro and collaborators, plus the four possible SNe by either catalog (2 out of these 12 sources are not in the plot because their \emph{k}-band magnitude is missing in the catalog). The plot shows three new objects which, according to their stellar-like color and their position in the diagram, turn out to be QSO-like AGNs; they are marked by a box. $K_s$ magnitudes are by \citet{McCracken}.}\label{fig:rzk}
   \end{figure*}
To investigate the nature of our AGN candidates and, in particular, of the unclassified sources, we made use of a color-color diagram, \emph{r-z vs z-k}, as proposed by \citet{Nakos}. The choice of the colors is suitable for QSO identification since their SEDs are characterized by an excess of emission in the \emph{k} band and so, on a $z-k$ axis, they are redder than stars. Objects, depending on their nature, occupy two distinct loci on such a diagram, corresponding to a rather sharp stellar sequence and a more scattered region where galaxies lie. 
In order to obtain the diagram we made use of data from two additional COSMOS catalogs:
\begin{itemize}
\item the COSMOS ACS catalog \citep{Koekemoer, Scoville} from the Hubble Space Telescope, constructed from 575 ACS pointings; the catalog provides a more reliable stellarity index than the one derived from ground-based observations to distinguish extended (e.g., bright galaxies) from unresolved (faint galaxies, stars and QSOs) sources;
\item the COSMOS Intermediate and Broad Band Photometry catalog, with a limiting magnitude \emph{r}(AB) $\gtrsim 29$ mag; it provides the magnitudes in the $r$, $z$, and $k$ bands. The catalog includes photometry in all 25 optical/NIR narrow-, intermediate-, and broadband filters from \emph{u} to $K_s$. The photometry is computed at the position of the $i^*$-band image using \emph{SExtractor} in dual mode. The catalog supersedes the one by \citet{Capak}, with improved source detection and photometry extracted in $3\arcsec$ diameter apertures.
\end{itemize}
The \emph{r-z vs z-k} diagram is shown in Fig. \ref{fig:rzk}. With the help of an indicator of the morphology of the sources, the diagram allows identification of QSO-like objects, i.e., compact sources showing non-stellar colors. The larger dots represent all the AGN candidates (not including the confirmed or possible SNe) in our list for which the \emph{r}, \emph{z}, and \emph{k} magnitudes and the stellarity index are available (67 out of 71 sources); their color is defined by the stellarity index. The crosses represent all the AGNs already confirmed so far mainly through X-ray validation, while the triangles are for the SNe. It is apparent that the sources in the plot define two distinct regions corresponding to stellar-like and extended objects. The diagram allows identification of three new QSO-like AGNs (red dots marked by a box and lying in the extended object locus); we classify as QSO-like all the sources lying on the galaxy locus and with a stellarity index $\geq 0.8$. All the SNe are characterized by colors typical of galaxies and they all lie in the galactic area of the diagram. We point out that, if no additional X-ray or spectroscopic information about the nature of our sample of optically variable sources were available, 54 out of 83 sources would be classified as QSOs on the basis of variability, color, and stellarity; as a consequence, we confirm the nature of $65\%$ of the sources in our sample on the basis of the sole \emph{r-z vs z-k} diagram.
We excluded the five sources with $z-k < 0.05$ on the basis of the star/AGN separation criteria by \citet{Nakos}, converted to our AB magnitude system.

While the availability of NIR data allows us to separate AGNs from galaxies better than the traditional optical colors (e.g., \emph{U-B vs B-V}), selecting sources redder than the galaxy locus in Fig. \ref{fig:rzk}, we point out that a sizable fraction of AGNs have colors consistent with normal galaxies and would have been lost without the variability criterion. Furthermore, the lost fraction is larger for fainter AGNs (LLAGNs), where the host galaxy contamination is more severe.

\section{Discussion and conclusions}
\label{section:discussion}
In the present work we derived a sample of optically variable sources and validated the nature of $94\%$ of them, as summarized below. Sixty-six objects in the sample turned out to be AGNs, proving the strength of optical variability as an AGN selection technique. This number corresponds to a density of 86 AGNs per square degree. None of the nine sources that we assumed to be SNe has a counterpart in the X-ray catalogs, and no evidence against the assumption that they are SNe was found; as a consequence, we confirm their classification as SNe, and classify the remaining three as possible SNe.

The purity\footnote{We define the purity as the number of confirmed AGNs divided by the number of AGN candidates (assuming that no information about the SNe is available \emph{a priori}, all the sources in the secure sample are AGN candidates). Conversely, the contamination is defined as the number of confirmed non-AGNs divided by the number of AGN candidates. Purity and contamination are, of course, complementary.} of our sample of optically variable sources is $80\%$, and it rises to 93\% if we do not include the confirmed or possible SNe in our sample of AGN candidates; the contamination of the sample ranges from 14\% to 20\% depending on whether we exclude or not the five non-classified sources from the contaminants. 

Most ($66\%$) of the sources in the secure sample have been confirmed by means of spectroscopic/SED classification, X/O, and color-color diagrams as well. 
Of the remaining five non-classified sources, one has an optical counterpart in various optical catalogs \citep{Capak, Ilbert, Koekemoer, Scoville}; the remaining four objects do not have any counterparts within a $1\arcsec$ radius, but we found that one of them has a counterpart within $1.08\arcsec$ in all the just mentioned catalogs plus the \emph{XMM} catalog. The source is spectroscopically classified as BLAGN in the \emph{XMM} catalog, its X-ray luminosity is $L_{\scriptscriptstyle X}>10^{42}$ erg s$^{-1}$, and its X/O (both soft and hard) is in the range [-1;1]; with respect to the \emph{r-z vs z-k} diagram, the object is QSO-like, so we can state it is an AGN after all the analyzed diagnostics. If we include this source in the confirmed AGNs, the purity of our sample rises to 81\%.

The nature of the other three variable sources is still unknown: each of them is detected in 25 out of the 27 epochs constituting our dataset; from Table \ref{tab:secure_sample} we can see that they all are rather faint (\emph{r}(AB) $> 21.7$ mag) and rather compact (stellarity index $> 0.7$). Two of them fall in the outer region of the \emph{Chandra}-COSMOS field, where the sensitivity is lower: the non-uniform depth of the X-ray catalog is a possible explanation for the lack of an X-ray counterpart. We compared the \emph{VST}-COSMOS images to those from \emph{HST} and \emph{CFHT}, and did not find any of the three objects although, given the depth of the observations, they should have been easily observed. Thus we conclude that they are real variable sources: either AGNs with weak X-ray emission or some other class of transient objects. 

In Table \ref{tab:validation} we list the number of sources confirmed by each diagnostic or combination of them; the additional confirmed AGN that we just mentioned is included. A complete list of the 83 sources in the secure sample is provided in Table \ref{tab:secure_sample}: for each object we give the coordinates from the stacked image, the average magnitude, the light curve r.m.s. (ltc r.m.s.), the stellarity index from the stacked image, the quality label that we attributed to the source, the significance, the spectroscopic redshift, and the source classification index, providing information about each diagnostic used to confirm each source.

\begin{table}[tb]
\caption{Confirmed sources. We list the number of objects confirmed by each diagnostic (lines 4 to 6), and also the number of sources (when it is not 0) confirmed by each combination of indicators (lines 7 to 10).} 
\label{tab:validation}      
\centering
\begin{tabular}{c c}
\hline\hline
confirmed sources (either AGNs or SNe) & 79 (95\% of 83)\\
\hline confirmed AGNs & 67 (81\% of 83)\\
confirmed SNe & 12 (14\% of 83)\\
\hline
\hline
\ spectroscopic/SED validation (S) & 63\\
X/O validation (X/O) & 64\\
color-color diagram validation (C)& 55\\
\hline
\hline
\ S+X/O+C validation & 51\\
S+X/O validation & 12\\
X/O+C validation & 1\\
only C validation & 3\\
\hline
\hline classified sources & \\
with no X-ray counterpart & 15\\
\hline
\end{tabular}
\end{table}

\begin{figure}[tb]
           {\includegraphics[width=\hsize]{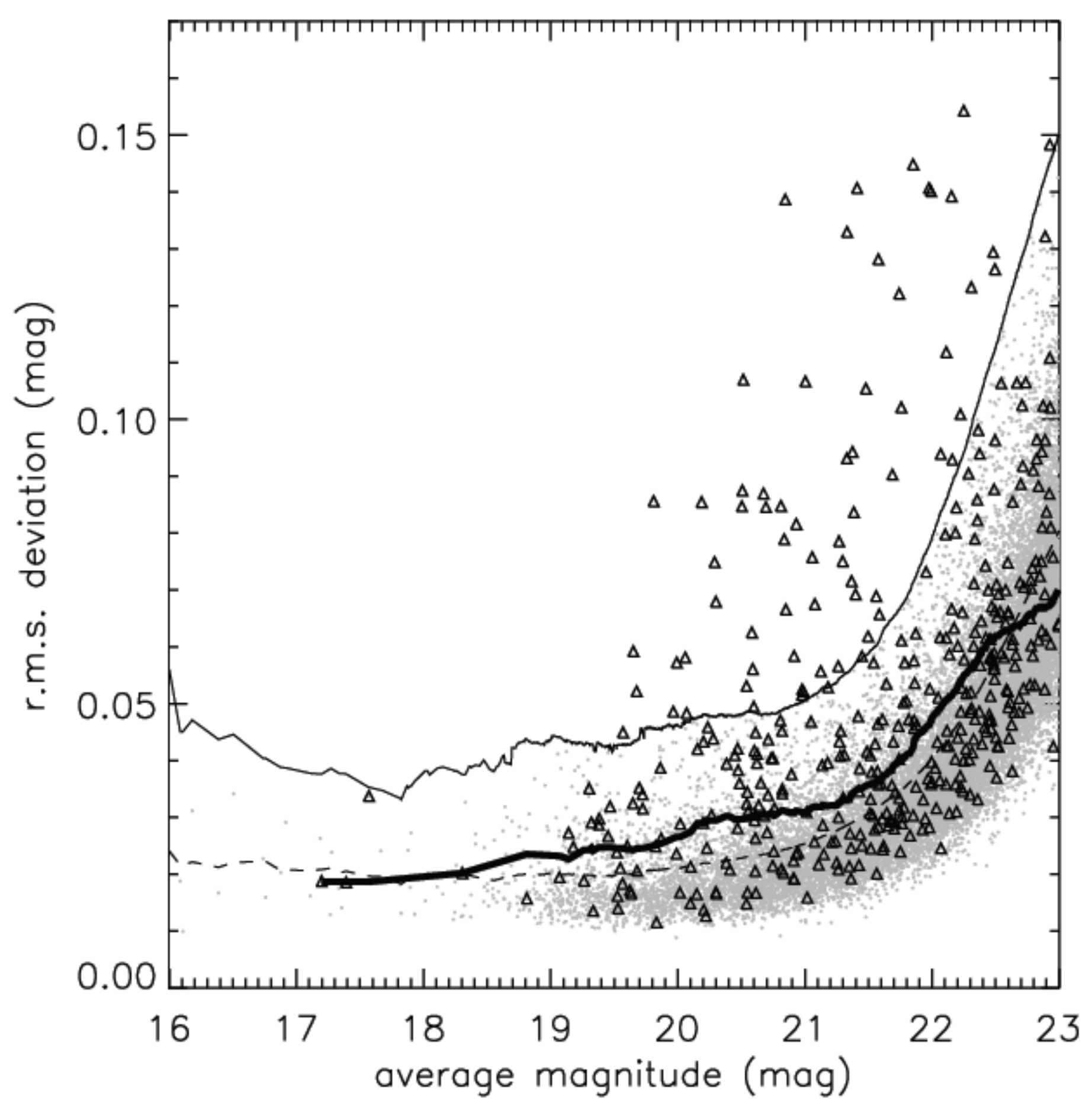}}
\caption{\footnotesize r.m.s. deviation from average magnitude as a function of average magnitude for the X-ray emitters that are confirmed AGNs (triangles). The grey dots represent all the non-variable sources in the \emph{VST} complete sample. The running average of the r.m.s. deviation of the complete sample (thin solid line) and of the subsample of X-ray emitters (thick solid line), plus the variability threshold (dashed line) are also shown. $85\%$ of the X-ray sources fall below the variability threshold.}
         \label{fig:stdev_vs_mag2}
   \end{figure}

The 63 X-ray emitting sources that are confirmed AGNs after our variability analysis correspond to $15\%$ of the X-ray emitters in the X-ray sample that are also AGNs, confirmed by means of spectroscopic classification and/or X-ray properties. This percentage defines the completeness\footnote{We define the completeness as the number of confirmed AGNs divided by the number of AGNs that were known \emph{a priori}.} of our secure sample with respect to the X-ray sources that are confirmed AGNs. We also computed the completeness in three magnitude bins of the same size from \emph{r}(AB) $=20$ mag to \emph{r}(AB) $=23$ mag: it is $26\%$ in the $20-21$ mag bin, then $23\%$ in the following bin, and it drops to $5\%$ for fainter magnitudes. In an attempt to explain the low completeness, we show in Fig. \ref{fig:stdev_vs_mag2} (which is similar to Fig. \ref{fig:stdev_vs_mag}) the location of all the X-ray sources with \emph{VST} counterparts and that are also confirmed AGNs. It is apparent that most of them are below the variability threshold, but on average they have larger r.m.s. than the rest of the population; this means that they are often optically variable, although we cannot detect their variability with the current photometric accuracy. The high average optical variability of the X-ray sources was proved by means of a Kolmogorov-Smirnov (K-S), test where we compared the r.m.s. of the sample of the \emph{VST} non-variable sources to the r.m.s. of the X-ray emitters with \emph{VST} counterparts and with an X-ray luminosity $L_{\scriptscriptstyle X} > 10^{42}$ erg s$^{-1}$; the test returned a probability $P\approx 10^{-7}$ that the two datasets were drawn from the same distribution. 
As a further test, we made this same comparison in the above mentioned three magnitude bins: we noticed that the X-ray sample is always characterized by an average optical variability higher than the average r.m.s. of the optical source population.
All the performed tests show that the optical variability of the X-ray sources can be overshadowed by the large uncertainties, especially when dealing with faint objects, and this explains the low completeness with respect to the X-ray sample; we are therefore confident that a higher photometric accuracy would return a higher completeness. This could also be achieved by lowering the variability threshold, although it would be at the expense of the purity of the selected sample.
The probability of the K-S test changes if we restrict the comparison to the subsamples of Type 1/Type 2 AGNs or, similarly, to the soft/hard\footnote{We define as ``soft'' an X-ray source having a hardness ratio (HR) $< -0.2$, hence a HR $>-0.2$ characterizes hard X-ray sources (see, e.g., \citealt[][and references therein]{Brusa}). The HR is defined as $(H-S)/(H+S)$, where \emph{H} and \emph{S} are the hard- and soft-band counts, respectively. A measure of the HR is found in the X-ray catalogs for 438 out of the 548 X-ray COSMOS sources with a \emph{VST} counterpart; when both \emph{Chandra} \citep{Elvis} and \emph{XMM} values were available, we adopted the latest measure, i.e., the one from \emph{XMM}.} X-ray sources (with $L_{\scriptscriptstyle X} > 10^{42}$ erg s$^{-1}$): while it is on the order of $P=10^{-4}-10^{-5}$ for Type 1 and soft X-ray AGNs, the variability is less evident for both Type 2 and hard X-ray sources, with $P=10^{-2}$. This suggests that Type 2/hard AGNs still present variability in excess with respect to the quiescent population, especially considering that they are on average fainter than the unobscured AGN population and thus their variability is harder to detect.

The completeness with respect to Type 1 AGNs is $26\%$, while it drops to $6\%$ for Type 2 AGNs. We attribute this difference to the lower average flux of the Type 2 sample, as suggested by the dependence of the completeness level on the source magnitude, as described above, although we cannot exclude that some intrinsic difference (e.g., variability dilution by the obscuring/reflective material) is present as well.

In \citet{Trevese} an optical variability analysis was performed over a set of eight epochs of the CDFS covering a $0.25$ square degree area, spanning about two years and imaged by the ESO/MPI 2.2 m telescope in La Silla, in the framework of the Southern inTermediate Redshift ESO Supernova Search (STRESS) survey. 
The analysis was limited to the 104 sources lying in the region covered by X-ray data. 
The purity of their sample is $\approx 60\%$, but this is a lower limit, because unconfirmed LLAGNs could be among the remaining $40\%$ of the candidates. 
The completeness with respect to the X-ray sources in the field (with known spectra and X-ray luminosity $L_{\scriptscriptstyle X}$ (2-8 keV) $> 10^{42}$ erg s$^{-1}$) is $44\%$; this suggests that a longer temporal baseline than ours (two years \emph{vs} five months) can lead to a higher completeness.
We tested this conclusion using the AGN structure function (e.g., \citealt{Bauer} and references therein) to estimate that increasing the baseline from six months to two years results in an increase of the intrinsic variability by $\sim 50\%$. Assuming the X-ray detected population to be representative of the whole population, we calculate that increasing their intrinsic variability by 50\%, and improving our photometric accuracy (\citealt{Trevese} have a slightly better accuracy than our early VST data) would bring $\sim 36\%\pm 3\%$ of the sources above our variability threshold; this is in agreement with the completeness of $44\%^{+6\%}_{-9\%}$ ($1\sigma$ binomial confidence limits) measured by \citet{Trevese}.
Moreover, in their work, as well as in ours, several variable extended objects are narrow emission line galaxies (NELGs), suggesting that they are low ionization narrow emission regions (LINERs) hosting LLAGNs. Variability studies thus provide samples of sources that are interesting for the study of AGN-starburst connection.

In \citet{Klesman&Sarajedini} the optical variability of a sample of IR- and X-ray-selected AGNs in the Great Observatories Origins Deep Survey (GOODS) South field was investigated: the sources in the sample have optical counterparts in the \emph{HST} ACS catalogs, and the analysis of five epochs over six months showed that $26\%$ of the X-ray emitters are optically variable; the chosen limiting magnitude is $V=27$, much fainter than ours, but the two baselines are comparable and the percentages of X-ray sources that also show optical variability are consistent: this suggests that the variability detection technique can be extended to fainter magnitudes (and presumably higher redshifts) provided we maintain a photometric accuracy below a few percentage points.\\

\citet{Villforth1} obtain similar results using the six epochs observed within the GOODS projects and spanning one year. They detect 139 AGNs down to $z \approx 25.5$ mag.Their method allows them to constrain with great accuracy the number of false detections, making use of a more refined statistical method that takes advantage of the accuracy of photometric error measurements in the \emph{HST} observations. In our case we can assess directly the number of false detections produced by our method taking advantage of the wealth of diagnostics available within the COSMOS field: as described above, only four sources lack any diagnostic to classify them as a SN or an AGN. Thus, the false detection rate is $< 5\%$, which drops to $\sim 1\%$ if we consider that three of these objects seem to be real transients based on the comparison with older \emph{HST} observations.  
A significant fraction of the sources identified by \citep{Villforth} are low-luminosity AGNs (based on SED fitting) without X-ray detections. This is at odds with our findings that the majority of our sources are X-ray emitters; however this difference is explained by the fact that our method is best suited to identifying bright AGN-dominated sources, i.e., QSO-like objects, while the use of \emph{HST} resolution allows Villforth et al. to better disentangle the emission of the central AGN from the host galaxy, probing lower luminosity sources.

\begin{figure}[tb]
           {\includegraphics[width=\hsize]{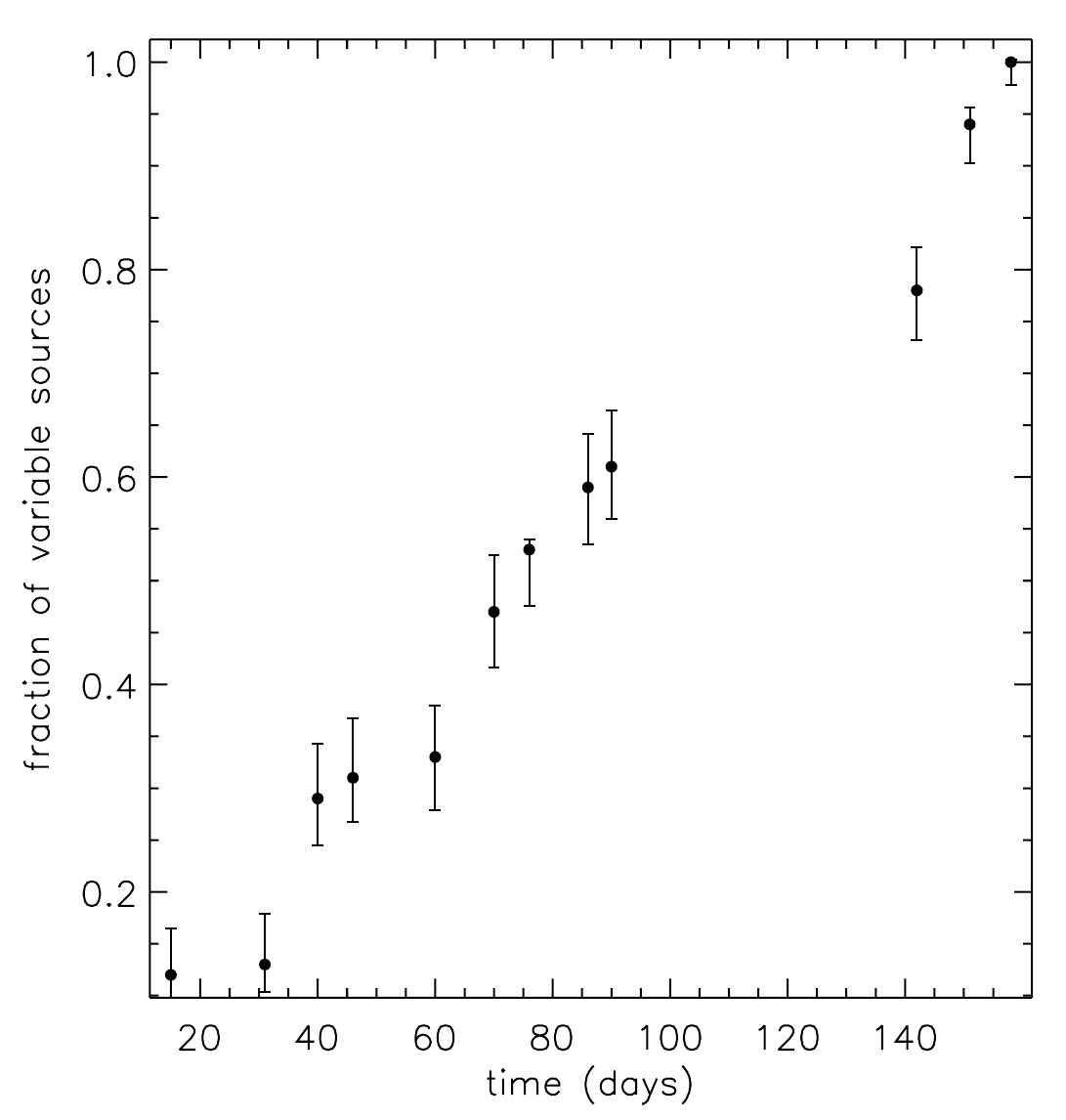}}
\caption{\footnotesize Fraction of sources in the secure sample retrieved for different baselines. Error bars are $1 \sigma$ binomial confidence limits.}
         \label{fig:var_sources_vs_time}
   \end{figure}

To test how our detection efficiency depends on the sampled timescales, we measured the number of variable sources recovered using different timescales. Figure \ref{fig:var_sources_vs_time} shows that the longer the timescale, the larger the fraction of variable sources.
This is expected on the basis of the known behavior of AGN variability, which exhibits a red noise power spectrum and an increasing structure function toward longer timescales \citep[e.g.,][and references therein]{deVries}, hence we also expect a substantial increase in the completeness with respect to the confirmed AGNs if observations are performed over a longer baseline. Longer timescales would also allow us to detect the variability of less variable galaxies, such as extended objects that may appear less variable because of the dilution caused by the light from the host galaxy; this would increase the completeness for faint AGNs. This is not yet possible with the present data, but it will certainly be possible with future observations. Moreover, if observations in this same field were repeated after one or more years, the probability of finding new SN candidates would not change, while the completeness of the AGN selection would certainly increase. 

To sum up, our results show how the selection of AGN candidates on the basis of their optical variability allows construction of robust AGN samples; this, especially when coupled with a higher photometric accuracy and a longer observing baseline, is encouraging in the framework of current and future wide-field surveys (e.g., DES, \emph{LSST}; see \citealt{Brandt}), where variability is important both for the discovery and the study of AGNs and other variable sources, confirming and strengthening the predictions made by \citet{Schmidt}. We refer to \cite{Falocco} for the analysis of optical variability in the CDFS, where additional data from IR observatories will be used to assess the quality of the selected sample of AGN candidates.

\begin{acknowledgements}
This work was supported by the PRIN-INAF "GALAXY EVOLUTION WITH THE \emph{VLT} SURVEYS TELESCOPE (VST)" (PI A. Grado).\\
Funding for SDSS-III has been provided by the Alfred P. Sloan Foundation, the Participating Institutions, the National Science Foundation, and the U.S. Department of Energy Office of Science. The SDSS-III web site is http://www.sdss3.org/. 
SDSS-III is managed by the Astrophysical Research Consortium for the Participating Institutions of the SDSS-III Collaboration including the University of Arizona, the Brazilian Participation Group, Brookhaven National Laboratory, University of Cambridge, Carnegie Mellon University, University of Florida, the French Participation Group, the German Participation Group, Harvard University, the Instituto de Astrofisica de Canarias, the Michigan State/Notre Dame/JINA Participation Group, Johns Hopkins University, Lawrence Berkeley National Laboratory, Max Planck Institute for Astrophysics, Max Planck Institute for Extraterrestrial Physics, New Mexico State University, New York University, Ohio State University, Pennsylvania State University, University of Portsmouth, Princeton University, the Spanish Participation Group, University of Tokyo, University of Utah, Vanderbilt University, University of Virginia, University of Washington, and Yale University.\\
GP acknowledges support provided by the Millennium Institute of Astrophysics through grant IC120009 of the Programa Iniciativa Cientifica Milenio del Ministerio de Economia, Fomento y Turismo de Chile.\\
MV acknowledges funding from the Square Kilometre Array South Africa
project and from the South African National Research Foundation.\\
We thank Marcella Brusa, Francesca Civano and Stefano Marchesi for providing access to updated versions of the COSMOS catalogs. We also thank the anonymous referee for helpful comments and suggestions.
\end{acknowledgements}

\bibliographystyle{aa}
\bibliography{arxiv1412_1488}{}

\begin{thebibliography}{65}
\expandafter\ifx\csname natexlab\endcsname\relax\def\natexlab#1{#1}\fi

\bibitem[{{Aihara} {et~al.}(2011){Aihara}, {Allende Prieto}, {An}, {Anderson},
  {Aubourg}, {Balbinot}, {Beers}, {Berlind}, {Bickerton}, {Bizyaev}, {Blanton},
  {Bochanski}, {Bolton}, {Bovy}, {Brandt}, {Brinkmann}, {Brown}, {Brownstein},
  {Busca}, {Campbell}, {Carr}, {Chen}, {Chiappini}, {Comparat}, {Connolly},
  {Cortes}, {Croft}, {Cuesta}, {da Costa}, {Davenport}, {Dawson}, {Dhital},
  {Ealet}, {Ebelke}, {Edmondson}, {Eisenstein}, {Escoffier}, {Esposito},
  {Evans}, {Fan}, {Femen{\'{\i}}a Castell{\'a}}, {Font-Ribera}, {Frinchaboy},
  {Ge}, {Gillespie}, {Gilmore}, {Gonz{\'a}lez Hern{\'a}ndez}, {Gott}, {Gould},
  {Grebel}, {Gunn}, {Hamilton}, {Harding}, {Harris}, {Hawley}, {Hearty}, {Ho},
  {Hogg}, {Holtzman}, {Honscheid}, {Inada}, {Ivans}, {Jiang}, {Johnson},
  {Jordan}, {Jordan}, {Kazin}, {Kirkby}, {Klaene}, {Knapp}, {Kneib},
  {Kochanek}, {Koesterke}, {Kollmeier}, {Kron}, {Lampeitl}, {Lang}, {Le Goff},
  {Lee}, {Lin}, {Long}, {Loomis}, {Lucatello}, {Lundgren}, {Lupton}, {Ma},
  {MacDonald}, {Mahadevan}, {Maia}, {Makler}, {Malanushenko}, {Malanushenko},
  {Mandelbaum}, {Maraston}, {Margala}, {Masters}, {McBride}, {McGehee},
  {McGreer}, {M{\'e}nard}, {Miralda-Escud{\'e}}, {Morrison}, {Mullally},
  {Muna}, {Munn}, {Murayama}, {Myers}, {Naugle}, {Neto}, {Nguyen}, {Nichol},
  {O'Connell}, {Ogando}, {Olmstead}, {Oravetz}, {Padmanabhan},
  {Palanque-Delabrouille}, {Pan}, {Pandey}, {P{\^a}ris}, {Percival},
  {Petitjean}, {Pfaffenberger}, {Pforr}, {Phleps}, {Pichon}, {Pieri}, {Prada},
  {Price-Whelan}, {Raddick}, {Ramos}, {Reyl{\'e}}, {Rich}, {Richards}, {Rix},
  {Robin}, {Rocha-Pinto}, {Rockosi}, {Roe}, {Rollinde}, {Ross}, {Ross},
  {Rossetto}, {S{\'a}nchez}, {Sayres}, {Schlegel}, {Schlesinger}, {Schmidt},
  {Schneider}, {Sheldon}, {Shu}, {Simmerer}, {Simmons}, {Sivarani}, {Snedden},
  {Sobeck}, {Steinmetz}, {Strauss}, {Szalay}, {Tanaka}, {Thakar}, {Thomas},
  {Tinker}, {Tofflemire}, {Tojeiro}, {Tremonti}, {Vandenberg}, {Vargas
  Maga{\~n}a}, {Verde}, {Vogt}, {Wake}, {Wang}, {Weaver}, {Weinberg}, {White},
  {White}, {Yanny}, {Yasuda}, {Yeche}, \& {Zehavi}}]{Aihara}
{Aihara}, H., {Allende Prieto}, C., {An}, D., {et~al.} 2011, \apjs, 193, 29

\bibitem[{{Albert} {et~al.}(2007){Albert}, {Aliu}, {Anderhub}, {Antoranz},
  {Armada}, {Baixeras}, {Barrio}, {Bartko}, {Bastieri}, {Becker}, {Bednarek},
  {Berger}, {Bigongiari}, {Biland}, {Bock}, {Bordas}, {Bosch-Ramon}, {Bretz},
  {Britvitch}, {Camara}, {Carmona}, {Chilingarian}, {Coarasa}, {Commichau},
  {Contreras}, {Cortina}, {Costado}, {Curtef}, {Danielyan}, {Dazzi}, {De
  Angelis}, {Delgado}, {de los Reyes}, {De Lotto}, {Domingo-Santamar{\'{\i}}a},
  {Dorner}, {Doro}, {Errando}, {Fagiolini}, {Ferenc}, {Fern{\'a}ndez}, {Firpo},
  {Flix}, {Fonseca}, {Font}, {Fuchs}, {Galante}, {Garc{\'{\i}}a-L{\'o}pez},
  {Garczarczyk}, {Gaug}, {Giller}, {Goebel}, {Hakobyan}, {Hayashida},
  {Hengstebeck}, {Herrero}, {H{\"o}hne}, {Hose}, {Hrupec}, {Hsu}, {Jacon},
  {Jogler}, {Kosyra}, {Kranich}, {Kritzer}, {Laille}, {Lindfors}, {Lombardi},
  {Longo}, {L{\'o}pez}, {L{\'o}pez}, {Lorenz}, {Majumdar}, {Maneva},
  {Mannheim}, {Mansutti}, {Mariotti}, {Mart{\'{\i}}nez}, {Mazin}, {Merck},
  {Meucci}, {Meyer}, {Miranda}, {Mirzoyan}, {Mizobuchi}, {Moralejo}, {Nieto},
  {Nilsson}, {Ninkovic}, {O{\~n}a-Wilhelmi}, {Otte}, {Oya}, {Paneque},
  {Panniello}, {Paoletti}, {Paredes}, {Pasanen}, {Pascoli}, {Pauss}, {Pegna},
  {Persic}, {Peruzzo}, {Piccioli}, {Prandini}, {Puchades}, {Raymers}, {Rhode},
  {Rib{\'o}}, {Rico}, {Rissi}, {Robert}, {R{\"u}gamer}, {Saggion}, {Saito},
  {S{\'a}nchez}, {Sartori}, {Scalzotto}, {Scapin}, {Schmitt}, {Schweizer},
  {Shayduk}, {Shinozaki}, {Shore}, {Sidro}, {Sillanp{\"a}{\"a}}, {Sobczynska},
  {Stamerra}, {Stark}, {Takalo}, {Tavecchio}, {Temnikov}, {Tescaro}, {Teshima},
  {Torres}, {Turini}, {Vankov}, {Vitale}, {Wagner}, {Wibig}, {Wittek},
  {Zandanel}, {Zanin}, \& {Zapatero}}]{Albert}
{Albert}, J., {Aliu}, E., {Anderhub}, H., {et~al.} 2007, \apj, 669, 862

\bibitem[{{Andersen} {et~al.}(1995){Andersen}, {Freyhammer}, \& {Storm}}]{IC}
{Andersen}, M.~I., {Freyhammer}, L., \& {Storm}, J. 1995, in European Southern
  Observatory Conference and Workshop Proceedings, Vol.~53, Calibrating and
  Understanding HST and ESO Instruments, ed. P.~{Benvenuti}, 87

\bibitem[{{Aretxaga} \& {Terlevich}(1994)}]{Aretxaga&Terlevich}
{Aretxaga}, I. \& {Terlevich}, R. 1994, \mnras, 269, 462

\bibitem[{{Baldwin} {et~al.}(1981){Baldwin}, {Phillips}, \& {Terlevich}}]{BPT}
{Baldwin}, J.~A., {Phillips}, M.~M., \& {Terlevich}, R. 1981, \pasp, 93, 5

\bibitem[{{Barr} \& {Mushotzky}(1986)}]{Barr&Mushotzky}
{Barr}, P. \& {Mushotzky}, R.~F. 1986, \nat, 320, 421

\bibitem[{{Bauer} {et~al.}(2009){Bauer}, {Baltay}, {Coppi}, {Ellman}, {Jerke},
  {Rabinowitz}, \& {Scalzo}}]{Bauer}
{Bauer}, A., {Baltay}, C., {Coppi}, P., {et~al.} 2009, \apj, 696, 1241

\bibitem[{{Bershady} {et~al.}(1998){Bershady}, {Trevese}, \& {Kron}}]{Bershady}
{Bershady}, M.~A., {Trevese}, D., \& {Kron}, R.~G. 1998, \apj, 496, 103

\bibitem[{{Bertin}(2006)}]{scamp}
{Bertin}, E. 2006, in Astronomical Society of the Pacific Conference Series,
  Vol. 351, Astronomical Data Analysis Software and Systems XV, ed.
  C.~{Gabriel}, C.~{Arviset}, D.~{Ponz}, \& S.~{Enrique}, 112

\bibitem[{{Bertin} \& {Arnouts}(1996)}]{Bertin}
{Bertin}, E. \& {Arnouts}, S. 1996, \aaps, 117, 393

\bibitem[{{Bertin} {et~al.}(2002){Bertin}, {Mellier}, {Radovich}, {Missonnier},
  {Didelon}, \& {Morin}}]{swarp}
{Bertin}, E., {Mellier}, Y., {Radovich}, M., {et~al.} 2002, in Astronomical
  Society of the Pacific Conference Series, Vol. 281, Astronomical Data
  Analysis Software and Systems XI, ed. D.~A. {Bohlender}, D.~{Durand}, \&
  T.~H. {Handley}, 228

\bibitem[{{Botticella} {et~al.}(2013){Botticella}, {Cappellaro}, {Pignata},
  {Baruffolo}, {Benetti}, {Bufano}, {Capaccioli}, {Cascone}, {Covone}, {Della
  Valle}, {Grado}, {Greggio}, {Limatola}, {Paolillo}, {Pastorello},
  {Tomasella}, {Turatto}, \& {Vaccari}}]{Botticella}
{Botticella}, M.~T., {Cappellaro}, E., {Pignata}, G., {et~al.} 2013, The
  Messenger, 151, 29

\bibitem[{{Boutsia} {et~al.}(2009){Boutsia}, {Leibundgut}, {Trevese}, \&
  {Vagnetti}}]{Boutsia}
{Boutsia}, K., {Leibundgut}, B., {Trevese}, D., \& {Vagnetti}, F. 2009, \aap,
  497, 81

\bibitem[{{Brandt}(et al. 2002)}]{Brandt}
{Brandt}, W.~N. et al. 2002, \emph{Active Galactic Nuclei,} in LSST Science
  Book, 293, 345

\bibitem[{{Brandt} \& {Hasinger}(2005)}]{Brandt&Hasinger}
{Brandt}, W.~N. \& {Hasinger}, G. 2005, \araa, 43, 827

\bibitem[{{Brusa} {et~al.}(2010){Brusa}, {Civano}, {Comastri}, {Miyaji},
  {Salvato}, {Zamorani}, {Cappelluti}, {Fiore}, {Hasinger}, {Mainieri},
  {Merloni}, {Bongiorno}, {Capak}, {Elvis}, {Gilli}, {Hao}, {Jahnke},
  {Koekemoer}, {Ilbert}, {Le Floc'h}, {Lusso}, {Mignoli}, {Schinnerer},
  {Silverman}, {Treister}, {Trump}, {Vignali}, {Zamojski}, {Aldcroft},
  {Aussel}, {Bardelli}, {Bolzonella}, {Cappi}, {Caputi}, {Contini},
  {Finoguenov}, {Fruscione}, {Garilli}, {Impey}, {Iovino}, {Iwasawa},
  {Kampczyk}, {Kartaltepe}, {Kneib}, {Knobel}, {Kovac}, {Lamareille},
  {Leborgne}, {Le Brun}, {Le Fevre}, {Lilly}, {Maier}, {McCracken}, {Pello},
  {Peng}, {Perez-Montero}, {de Ravel}, {Sanders}, {Scodeggio}, {Scoville},
  {Tanaka}, {Taniguchi}, {Tasca}, {de la Torre}, {Tresse}, {Vergani}, \&
  {Zucca}}]{Brusa}
{Brusa}, M., {Civano}, F., {Comastri}, A., {et~al.} 2010, \apj, 716, 348

\bibitem[{{Capaccioli} \& {Schipani}(2011)}]{VST}
{Capaccioli}, M. \& {Schipani}, P. 2011, The Messenger, 146, 2

\bibitem[{{Capak} {et~al.}(2007){Capak}, {Aussel}, {Ajiki}, {McCracken},
  {Mobasher}, {Scoville}, {Shopbell}, {Taniguchi}, {Thompson}, {Tribiano},
  {Sasaki}, {Blain}, {Brusa}, {Carilli}, {Comastri}, {Carollo}, {Cassata},
  {Colbert}, {Ellis}, {Elvis}, {Giavalisco}, {Green}, {Guzzo}, {Hasinger},
  {Ilbert}, {Impey}, {Jahnke}, {Kartaltepe}, {Kneib}, {Koda}, {Koekemoer},
  {Komiyama}, {Leauthaud}, {Le Fevre}, {Lilly}, {Liu}, {Massey}, {Miyazaki},
  {Murayama}, {Nagao}, {Peacock}, {Pickles}, {Porciani}, {Renzini}, {Rhodes},
  {Rich}, {Salvato}, {Sanders}, {Scarlata}, {Schiminovich}, {Schinnerer},
  {Scodeggio}, {Sheth}, {Shioya}, {Tasca}, {Taylor}, {Yan}, \&
  {Zamorani}}]{Capak}
{Capak}, P., {Aussel}, H., {Ajiki}, M., {et~al.} 2007, \apjs, 172, 99

\bibitem[{{Cappellaro}(et al. in preparation)}]{Cappellaro}
{Cappellaro}, E. et al. in preparation

\bibitem[{{Civano} {et~al.}(2012){Civano}, {Elvis}, {Lanzuisi}, {Aldcroft},
  {Trichas}, {Bongiorno}, {Brusa}, {Blecha}, {Comastri}, {Loeb}, {Salvato},
  {Fruscione}, {Koekemoer}, {Komossa}, {Gilli}, {Mainieri}, {Piconcelli}, \&
  {Vignali}}]{Civano}
{Civano}, F., {Elvis}, M., {Lanzuisi}, G., {et~al.} 2012, \apj, 752, 49

\bibitem[{{Comastri} {et~al.}(2010){Comastri}, {Iwasawa}, {Gilli}, {Vignali},
  {Ranalli}, {Matt}, \& {Fiore}}]{Comastri}
{Comastri}, A., {Iwasawa}, K., {Gilli}, R., {et~al.} 2010, \apj, 717, 787

\bibitem[{{Cristiani} {et~al.}(1996){Cristiani}, {Trentini}, {La Franca},
  {Aretxaga}, {Andreani}, {Vio}, \& {Gemmo}}]{Cristiani}
{Cristiani}, S., {Trentini}, S., {La Franca}, F., {et~al.} 1996, \aap, 306, 395

\bibitem[{{de Vries} {et~al.}(2005){de Vries}, {Becker}, {White}, \&
  {Loomis}}]{deVries}
{de Vries}, W.~H., {Becker}, R.~H., {White}, R.~L., \& {Loomis}, C. 2005, \aj,
  129, 615

\bibitem[{{Elvis} {et~al.}(2009){Elvis}, {Civano}, {Vignali}, {Puccetti},
  {Fiore}, {Cappelluti}, {Aldcroft}, {Fruscione}, {Zamorani}, {Comastri},
  {Brusa}, {Gilli}, {Miyaji}, {Damiani}, {Koekemoer}, {Finoguenov}, {Brunner},
  {Urry}, {Silverman}, {Mainieri}, {Hasinger}, {Griffiths}, {Carollo}, {Hao},
  {Guzzo}, {Blain}, {Calzetti}, {Carilli}, {Capak}, {Ettori}, {Fabbiano},
  {Impey}, {Lilly}, {Mobasher}, {Rich}, {Salvato}, {Sanders}, {Schinnerer},
  {Scoville}, {Shopbell}, {Taylor}, {Taniguchi}, \& {Volonteri}}]{Elvis}
{Elvis}, M., {Civano}, F., {Vignali}, C., {et~al.} 2009, \apjs, 184, 158

\bibitem[{{Falocco} {et~al.}(submitted){Falocco}, {Paolillo}, {Covone}, {De
  Cicco}, {Longo}, {Grado}, {Limatola}, {Vaccari}, {Botticella}, {Pignata},
  {Cappellaro}, {Trevese}, {Vagnetti}, {Salvato}, {Radovich}, {Hsu},
  {Capaccioli}, {Napolitano}, {Baruffolo}, \& {Cascone}}]{Falocco}
{Falocco}, S., {Paolillo}, M., {Covone}, G., {et~al.} submitted

\bibitem[{{Fan}(1999)}]{Fan}
{Fan}, X. 1999, \aj, 117, 2528

\bibitem[{{Ferrarese} \& {Merritt}(2000)}]{Ferrarese&Merritt}
{Ferrarese}, L. \& {Merritt}, D. 2000, \apjl, 539, L9

\bibitem[{{Fiore} {et~al.}(2009){Fiore}, {Puccetti}, {Brusa}, {Salvato},
  {Zamorani}, {Aldcroft}, {Aussel}, {Brunner}, {Capak}, {Cappelluti}, {Civano},
  {Comastri}, {Elvis}, {Feruglio}, {Finoguenov}, {Fruscione}, {Gilli},
  {Hasinger}, {Koekemoer}, {Kartaltepe}, {Ilbert}, {Impey}, {Le Floc'h},
  {Lilly}, {Mainieri}, {Martinez-Sansigre}, {McCracken}, {Menci}, {Merloni},
  {Miyaji}, {Sanders}, {Sargent}, {Schinnerer}, {Scoville}, {Silverman},
  {Smolcic}, {Steffen}, {Santini}, {Taniguchi}, {Thompson}, {Trump}, {Vignali},
  {Urry}, \& {Yan}}]{Fiore}
{Fiore}, F., {Puccetti}, S., {Brusa}, M., {et~al.} 2009, \apj, 693, 447

\bibitem[{{Gaskell} \& {Klimek}(2003)}]{Gaskell&Klimek}
{Gaskell}, C.~M. \& {Klimek}, E.~S. 2003, Astronomical and Astrophysical
  Transactions, 22, 661

\bibitem[{{Grado} {et~al.}(2012){Grado}, {Capaccioli}, {Limatola}, \&
  {Getman}}]{Grado}
{Grado}, A., {Capaccioli}, M., {Limatola}, L., \& {Getman}, F. 2012, Memorie
  della Societa Astronomica Italiana Supplementi, 19, 362

\bibitem[{{Huang} {et~al.}(2011){Huang}, {Radovich}, {Grado}, {Puddu},
  {Romano}, {Limatola}, \& {Fu}}]{Huang}
{Huang}, Z., {Radovich}, M., {Grado}, A., {et~al.} 2011, \aap, 529, A93

\bibitem[{{Ilbert} {et~al.}(2009){Ilbert}, {Capak}, {Salvato}, {Aussel},
  {McCracken}, {Sanders}, {Scoville}, {Kartaltepe}, {Arnouts}, {Le Floc'h},
  {Mobasher}, {Taniguchi}, {Lamareille}, {Leauthaud}, {Sasaki}, {Thompson},
  {Zamojski}, {Zamorani}, {Bardelli}, {Bolzonella}, {Bongiorno}, {Brusa},
  {Caputi}, {Carollo}, {Contini}, {Cook}, {Coppa}, {Cucciati}, {de la Torre},
  {de Ravel}, {Franzetti}, {Garilli}, {Hasinger}, {Iovino}, {Kampczyk},
  {Kneib}, {Knobel}, {Kovac}, {Le Borgne}, {Le Brun}, {F{\`e}vre}, {Lilly},
  {Looper}, {Maier}, {Mainieri}, {Mellier}, {Mignoli}, {Murayama}, {Pell{\`o}},
  {Peng}, {P{\'e}rez-Montero}, {Renzini}, {Ricciardelli}, {Schiminovich},
  {Scodeggio}, {Shioya}, {Silverman}, {Surace}, {Tanaka}, {Tasca}, {Tresse},
  {Vergani}, \& {Zucca}}]{Ilbert1}
{Ilbert}, O., {Capak}, P., {Salvato}, M., {et~al.} 2009, \apj, 690, 1236

\bibitem[{{Ilbert} {et~al.}(2008){Ilbert}, {Salvato}, {Capak}, {Le Floc'h},
  {Aussel}, {McCracken}, {Arnouts}, {Mobasher}, {Sanders}, {Scoville}, \&
  {Taniguchi}}]{Ilbert}
{Ilbert}, O., {Salvato}, M., {Capak}, P., {et~al.} 2008, in Astronomical
  Society of the Pacific Conference Series, Vol. 399, Panoramic Views of Galaxy
  Formation and Evolution, ed. T.~{Kodama}, T.~{Yamada}, \& K.~{Aoki}, 169

\bibitem[{{Klesman} \& {Sarajedini}(2007)}]{Klesman&Sarajedini}
{Klesman}, A. \& {Sarajedini}, V. 2007, \apj, 665, 225

\bibitem[{{Koekemoer} {et~al.}(2007){Koekemoer}, {Aussel}, {Calzetti}, {Capak},
  {Giavalisco}, {Kneib}, {Leauthaud}, {Le F{\`e}vre}, {McCracken}, {Massey},
  {Mobasher}, {Rhodes}, {Scoville}, \& {Shopbell}}]{Koekemoer}
{Koekemoer}, A.~M., {Aussel}, H., {Calzetti}, D., {et~al.} 2007, \apjs, 172,
  196

\bibitem[{{Kormendy} \& {Ho}(2013)}]{Kormendy&Ho}
{Kormendy}, J. \& {Ho}, L.~C. 2013, \araa, 51, 511

\bibitem[{{Kormendy} \& {Richstone}(1995)}]{Kormendy&Richstone}
{Kormendy}, J. \& {Richstone}, D. 1995, \araa, 33, 581

\bibitem[{{Krolik} {et~al.}(1991){Krolik}, {Horne}, {Kallman}, {Malkan},
  {Edelson}, \& {Kriss}}]{Krolik}
{Krolik}, J.~H., {Horne}, K., {Kallman}, T.~R., {et~al.} 1991, \apj, 371, 541

\bibitem[{{Kuijken}(2011)}]{Kuijken}
{Kuijken}, K. 2011, The Messenger, 146, 8

\bibitem[{{Lawrence} \& {Papadakis}(1993)}]{Lawrence&Papadakis}
{Lawrence}, A. \& {Papadakis}, I. 1993, \apjl, 414, L85

\bibitem[{{Lusso} {et~al.}(2010){Lusso}, {Comastri}, {Vignali}, {Zamorani},
  {Brusa}, {Gilli}, {Iwasawa}, {Salvato}, {Civano}, {Elvis}, {Merloni},
  {Bongiorno}, {Trump}, {Koekemoer}, {Schinnerer}, {Le Floc'h}, {Cappelluti},
  {Jahnke}, {Sargent}, {Silverman}, {Mainieri}, {Fiore}, {Bolzonella}, {Le
  F{\`e}vre}, {Garilli}, {Iovino}, {Kneib}, {Lamareille}, {Lilly}, {Mignoli},
  {Scodeggio}, \& {Vergani}}]{Lusso}
{Lusso}, E., {Comastri}, A., {Vignali}, C., {et~al.} 2010, \aap, 512, A34

\bibitem[{{Maccacaro} {et~al.}(1988){Maccacaro}, {Gioia}, {Wolter}, {Zamorani},
  \& {Stocke}}]{Maccacaro}
{Maccacaro}, T., {Gioia}, I.~M., {Wolter}, A., {Zamorani}, G., \& {Stocke},
  J.~T. 1988, \apj, 326, 680

\bibitem[{{Mainieri} {et~al.}(2002){Mainieri}, {Bergeron}, {Hasinger},
  {Lehmann}, {Rosati}, {Schmidt}, {Szokoly}, \& {Della Ceca}}]{Mainieri}
{Mainieri}, V., {Bergeron}, J., {Hasinger}, G., {et~al.} 2002, \aap, 393, 425

\bibitem[{{Malizia} {et~al.}(2014){Malizia}, {Molina}, {Bassani}, {Stephen},
  {Bazzano}, {Ubertini}, \& {Bird}}]{Malizia}
{Malizia}, A., {Molina}, M., {Bassani}, L., {et~al.} 2014, \apjl, 782, L25

\bibitem[{{McCracken} {et~al.}(2010){McCracken}, {Capak}, {Salvato}, {Aussel},
  {Thompson}, {Daddi}, {Sanders}, {Kneib}, {Willott}, {Mancini}, {Renzini},
  {Cook}, {Le F{\`e}vre}, {Ilbert}, {Kartaltepe}, {Koekemoer}, {Mellier},
  {Murayama}, {Scoville}, {Shioya}, \& {Tanaguchi}}]{McCracken}
{McCracken}, H.~J., {Capak}, P., {Salvato}, M., {et~al.} 2010, \apj, 708, 202

\bibitem[{{Muzzin} {et~al.}(2013){Muzzin}, {Marchesini}, {Stefanon}, {Franx},
  {Milvang-Jensen}, {Dunlop}, {Fynbo}, {Brammer}, {Labb{\'e}}, \& {van
  Dokkum}}]{Muzzin}
{Muzzin}, A., {Marchesini}, D., {Stefanon}, M., {et~al.} 2013, \apjs, 206, 8

\bibitem[{{Nakos} {et~al.}(2009){Nakos}, {Willis}, {Andreon}, {Surdej},
  {Riaud}, {Hatziminaoglou}, {Garcet}, {Alloin}, {Baes}, {Galaz}, {Pierre},
  {Quintana}, {Page}, {Tedds}, {Ceballos}, {Corral}, {Ebrero}, {Krumpe}, \&
  {Mateos}}]{Nakos}
{Nakos}, T., {Willis}, J.~P., {Andreon}, S., {et~al.} 2009, \aap, 494, 579

\bibitem[{{Paolillo} {et~al.}(2004){Paolillo}, {Schreier}, {Giacconi},
  {Koekemoer}, \& {Grogin}}]{Paolillo}
{Paolillo}, M., {Schreier}, E.~J., {Giacconi}, R., {Koekemoer}, A.~M., \&
  {Grogin}, N.~A. 2004, \apj, 611, 93

\bibitem[{{Pereyra} {et~al.}(2006){Pereyra}, {Vanden Berk}, {Turnshek},
  {Hillier}, {Wilhite}, {Kron}, {Schneider}, \& {Brinkmann}}]{Pereyra}
{Pereyra}, N.~A., {Vanden Berk}, D.~E., {Turnshek}, D.~A., {et~al.} 2006, \apj,
  642, 87

\bibitem[{{Richards} {et~al.}(2001){Richards}, {Fan}, {Schneider}, {Vanden
  Berk}, {Strauss}, {York}, {Anderson}, {Anderson}, {Annis}, {Bahcall},
  {Bernardi}, {Briggs}, {Brinkmann}, {Brunner}, {Burles}, {Carey}, {Castander},
  {Connolly}, {Crocker}, {Csabai}, {Doi}, {Finkbeiner}, {Friedman}, {Frieman},
  {Fukugita}, {Gunn}, {Hindsley}, {Ivezi{\'c}}, {Kent}, {Knapp}, {Lamb},
  {Leger}, {Long}, {Loveday}, {Lupton}, {McKay}, {Meiksin}, {Merrelli}, {Munn},
  {Newberg}, {Newcomb}, {Nichol}, {Owen}, {Pier}, {Pope}, {Richmond},
  {Rockosi}, {Schlegel}, {Siegmund}, {Smee}, {Snir}, {Stoughton}, {Stubbs},
  {SubbaRao}, {Szalay}, {Szokoly}, {Tremonti}, {Uomoto}, {Waddell}, {Yanny}, \&
  {Zheng}}]{Richards}
{Richards}, G.~T., {Fan}, X., {Schneider}, D.~P., {et~al.} 2001, \aj, 121, 2308

\bibitem[{{Salvato} {et~al.}(2009){Salvato}, {Hasinger}, {Ilbert}, {Zamorani},
  {Brusa}, {Scoville}, {Rau}, {Capak}, {Arnouts}, {Aussel}, {Bolzonella},
  {Buongiorno}, {Cappelluti}, {Caputi}, {Civano}, {Cook}, {Elvis}, {Gilli},
  {Jahnke}, {Kartaltepe}, {Impey}, {Lamareille}, {Le Floc'h}, {Lilly},
  {Mainieri}, {McCarthy}, {McCracken}, {Mignoli}, {Mobasher}, {Murayama},
  {Sasaki}, {Sanders}, {Schiminovich}, {Shioya}, {Shopbell}, {Silverman},
  {Smol{\v c}i{\'c}}, {Surace}, {Taniguchi}, {Thompson}, {Trump}, {Urry}, \&
  {Zamojski}}]{Salvato}
{Salvato}, M., {Hasinger}, G., {Ilbert}, O., {et~al.} 2009, \apj, 690, 1250

\bibitem[{{Salvato} {et~al.}(2011){Salvato}, {Ilbert}, {Hasinger}, {Rau},
  {Civano}, {Zamorani}, {Brusa}, {Elvis}, {Vignali}, {Aussel}, {Comastri},
  {Fiore}, {Le Floc'h}, {Mainieri}, {Bardelli}, {Bolzonella}, {Bongiorno},
  {Capak}, {Caputi}, {Cappelluti}, {Carollo}, {Contini}, {Garilli}, {Iovino},
  {Fotopoulou}, {Fruscione}, {Gilli}, {Halliday}, {Kneib}, {Kakazu},
  {Kartaltepe}, {Koekemoer}, {Kovac}, {Ideue}, {Ikeda}, {Impey}, {Le Fevre},
  {Lamareille}, {Lanzuisi}, {Le Borgne}, {Le Brun}, {Lilly}, {Maier},
  {Manohar}, {Masters}, {McCracken}, {Messias}, {Mignoli}, {Mobasher}, {Nagao},
  {Pello}, {Puccetti}, {Perez-Montero}, {Renzini}, {Sargent}, {Sanders},
  {Scodeggio}, {Scoville}, {Shopbell}, {Silvermann}, {Taniguchi}, {Tasca},
  {Tresse}, {Trump}, \& {Zucca}}]{Salvato1}
{Salvato}, M., {Ilbert}, O., {Hasinger}, G., {et~al.} 2011, \apj, 742, 61

\bibitem[{{Schmidt} {et~al.}(2010){Schmidt}, {Marshall}, {Rix}, {Jester},
  {Hennawi}, \& {Dobler}}]{Schmidt}
{Schmidt}, K.~B., {Marshall}, P.~J., {Rix}, H.-W., {et~al.} 2010, \apj, 714,
  1194

\bibitem[{{Scoville} {et~al.}(2007{\natexlab{a}}){Scoville}, {Abraham},
  {Aussel}, {Barnes}, {Benson}, {Blain}, {Calzetti}, {Comastri}, {Capak},
  {Carilli}, {Carlstrom}, {Carollo}, {Colbert}, {Daddi}, {Ellis}, {Elvis},
  {Ewald}, {Fall}, {Franceschini}, {Giavalisco}, {Green}, {Griffiths}, {Guzzo},
  {Hasinger}, {Impey}, {Kneib}, {Koda}, {Koekemoer}, {Lefevre}, {Lilly}, {Liu},
  {McCracken}, {Massey}, {Mellier}, {Miyazaki}, {Mobasher}, {Mould}, {Norman},
  {Refregier}, {Renzini}, {Rhodes}, {Rich}, {Sanders}, {Schiminovich},
  {Schinnerer}, {Scodeggio}, {Sheth}, {Shopbell}, {Taniguchi}, {Tyson}, {Urry},
  {Van Waerbeke}, {Vettolani}, {White}, \& {Yan}}]{Scoville}
{Scoville}, N., {Abraham}, R.~G., {Aussel}, H., {et~al.} 2007{\natexlab{a}},
  \apjs, 172, 38

\bibitem[{{Scoville} {et~al.}(2007{\natexlab{b}}){Scoville}, {Aussel}, {Brusa},
  {Capak}, {Carollo}, {Elvis}, {Giavalisco}, {Guzzo}, {Hasinger}, {Impey},
  {Kneib}, {LeFevre}, {Lilly}, {Mobasher}, {Renzini}, {Rich}, {Sanders},
  {Schinnerer}, {Schminovich}, {Shopbell}, {Taniguchi}, \& {Tyson}}]{Scoville1}
{Scoville}, N., {Aussel}, H., {Brusa}, M., {et~al.} 2007{\natexlab{b}}, \apjs,
  172, 1

\bibitem[{{Skrutskie} {et~al.}(2006){Skrutskie}, {Cutri}, {Stiening},
  {Weinberg}, {Schneider}, {Carpenter}, {Beichman}, {Capps}, {Chester},
  {Elias}, {Huchra}, {Liebert}, {Lonsdale}, {Monet}, {Price}, {Seitzer},
  {Jarrett}, {Kirkpatrick}, {Gizis}, {Howard}, {Evans}, {Fowler}, {Fullmer},
  {Hurt}, {Light}, {Kopan}, {Marsh}, {McCallon}, {Tam}, {Van Dyk}, \&
  {Wheelock}}]{2mass}
{Skrutskie}, M.~F., {Cutri}, R.~M., {Stiening}, R., {et~al.} 2006, \aj, 131,
  1163

\bibitem[{{Trevese} {et~al.}(2008){Trevese}, {Boutsia}, {Vagnetti},
  {Cappellaro}, \& {Puccetti}}]{Trevese}
{Trevese}, D., {Boutsia}, K., {Vagnetti}, F., {Cappellaro}, E., \& {Puccetti},
  S. 2008, \aap, 488, 73

\bibitem[{{Ulrich} {et~al.}(1997){Ulrich}, {Maraschi}, \& {Urry}}]{Ulrich}
{Ulrich}, M.-H., {Maraschi}, L., \& {Urry}, C.~M. 1997, \araa, 35, 445

\bibitem[{{Uttley}(2006)}]{Uttley}
{Uttley}, P. 2006, in Astronomical Society of the Pacific Conference Series,
  Vol. 360, Astronomical Society of the Pacific Conference Series, ed. C.~M.
  {Gaskell}, I.~M. {McHardy}, B.~M. {Peterson}, \& S.~G. {Sergeev}, 101

\bibitem[{{Vaccari}(et al. in preparation)}]{Vaccari}
{Vaccari}, M. et al. in preparation

\bibitem[{{Villforth} {et~al.}(2010){Villforth}, {Koekemoer}, \&
  {Grogin}}]{Villforth1}
{Villforth}, C., {Koekemoer}, A.~M., \& {Grogin}, N.~A. 2010, \apj, 723, 737

\bibitem[{{Villforth} {et~al.}(2012){Villforth}, {Sarajedini}, \&
  {Koekemoer}}]{Villforth}
{Villforth}, C., {Sarajedini}, V., \& {Koekemoer}, A. 2012, \mnras, 426, 360

\bibitem[{{Webb} \& {Malkan}(2000)}]{Webb&Malkan}
{Webb}, W. \& {Malkan}, M. 2000, \apj, 540, 652

\bibitem[{{Wood}(2011)}]{gam}
{Wood}, S.~N. 2011, {\emph{Fast stable restricted maximum likelihood and
  marginal likelihood estimation of semiparametric generalized linear models,}
  in Journal of the Royal Statistical Society (B), 73 (1), 3}

\bibitem[{{Xue} {et~al.}(2011){Xue}, {Luo}, {Brandt}, {Bauer}, {Lehmer},
  {Broos}, {Schneider}, {Alexander}, {Brusa}, {Comastri}, {Fabian}, {Gilli},
  {Hasinger}, {Hornschemeier}, {Koekemoer}, {Liu}, {Mainieri}, {Paolillo},
  {Rafferty}, {Rosati}, {Shemmer}, {Silverman}, {Smail}, {Tozzi}, \&
  {Vignali}}]{Xue}
{Xue}, Y.~Q., {Luo}, B., {Brandt}, W.~N., {et~al.} 2011, \apjs, 195, 10

\end{thebibliography}


\clearpage
\onecolumn
\begin{landscape}
\begin{longtable}{c c c c c c c c c c}
\caption{List of the 83 optically variable sources in the secure sample. Column meanings: (1): identification number; (2) and (3): right ascension and declination (J2000); (4): average \emph{VST r}(AB) magnitude; (5): r.m.s. of the light curve, $ltc\mbox{ }r.m.s.=\langle\sigma_i^{ltc}\rangle + 3\times\mbox{r.m.s.}_{\langle\sigma_i^{ltc}\rangle}$ (see Eq. \ref{eqn:var_threshold}); (6): \emph{SExtractor} stellarity index; (7): quality label; (8): significance (see Eq. \ref{eqn:significance}); (9): spectroscopic redshift by \citet{Civano} or, when not available, by \citet{Brusa}; (10): source classification. The classification index is the sum of different numbers corresponding to the following key:\\
1 = confirmed AGN through spectroscopy/SED; 2=confirmed AGN through X/O diagram; 4 = confirmed AGN through color \emph{vs} color diagram; 
0 = non-classified; 00 = non-classified, and no counterpart in the other COSMOS catalogs inspected; -1 = SN; -2 = possible SN.}\label{tab:secure_sample}\\
\hline source ID & RA J2000 (hms) & Dec J2000 (dms) & avg $r(AB)$ mag (mag) & ltc r.m.s. (mag) & stellarity & quality label & $\sigma^*$ & spectroscopic redshift & classification\\
(1) & (2) & (3) & (4) & (5) & (6) & (7) & (8) & (9) & (10)\\
\endfirsthead
\caption{Continued.} \\
\hline source ID & RA J2000 (hms) & Dec J2000 (dms) & avg $r(AB)$ mag (mag) & ltc r.m.s. (mag) & stellarity & quality label & $\sigma^*$ & spectroscopic redshift & classification\\
(1) & (2) & (3) & (4) & (5) & (6) & (7) & (8) & (9) & (10)\\
\hline
\endhead
\hline
\endfoot
\hline
\ 1 & 10:01:47.4 & +01:41:44.5 & 22.29 & 0.10 & 0.97 & 1 & 3.5 & - & 4\\
\ 2 & 09:58:49.5 & +01:42:20.7 & 20.69 & 0.10 & 0.03 & 1 & 9.8 & - & -1\\
\ 3 & 10:02:12.1 & +01:42:32.5 & 20.30 & 0.07 & 0.76 & 1 & 5.4 & 0.369 & 7\\
\ 4 & 09:59:13.9 & +01:42:33.5 & 22.11 & 0.14 & 0.98 & 1 & 6.5 & - & 4\\
\ 5 & 09:59:07.1 & +01:42:56.3 & 20.90 & 0.08 & 0.03 & 1 & 7.2 & - & -1\\
\ 6 & 09:58:45.0 & +01:43:08.9 & 20.19 & 0.09 & 0.90 & 1 & 7.4 & 1.337 & 7\\
\ 7 & 09:59:58.0 & +01:43:27.6 & 20.51 & 0.09 & 0.93 & 1 & 7.7 & 1.619 & 7\\
\ 8 & 10:00:12.1 & +01:44:40.0 & 22.22 & 0.10 & 0.72 & 1 & 3.6 & 1.148 & 3\\
\ 9 & 10:02:08.6 & +01:45:53.6 & 21.50 & 0.06 & 0.94 & 1 & 3.3 & 2.204 & 7\\
\ 10 & 10:02:23.0 & +01:47:14.9 & 21.45 & 0.06 & 0.89 & 1 & 3.0 & 1.246 & 7\\
\ 11 & 10:00:31.6 & +01:47:57.7 & 20.98 & 0.05 & 0.93 & 1 & 3.3 & 1.679 & 7\\
\ 12 & 09:58:34.3 & +01:51:36.6 & 22.91 & 0.72 & 0.07 & 1 & 28.5 & - & -1\\
\ 13 & 10:00:50.0 & +01:52:31.5 & 20.59 & 0.06 & 0.91 & 1 & 3.9 & 1.156 & 7\\
\ 14 & 10:00:33.4 & +01:52:37.1 & 20.69 & 0.08 & 0.90 & 1 & 7.4 & 0.832 & 7\\
\ 15 & 10:00:58.8 & +01:54:00.7 & 19.81 & 0.09 & 0.91 & 1 & 7.9 & 1.557 & 7\\
\ 16 & 10:01:36.2 & +01:54:43.0 & 21.27 & 0.08 & 0.95 & 1 & 5.8 & 2.281 & 7\\
\ 17 & 10:02:19.5 & +01:55:37.2 & 19.65& 0.06 & 0.86 & 1 & 5.1 & 1.509 & 7\\
\ 18 & 10:01:43.4 & +01:56:06.8 & 21.30 & 0.08 & 0.98 & 1 & 5.3 & 2.181 & 7\\
\ 19 & 10:01:24.7 & +01:57:38.8 & 21.40 & 0.07 & 0.95 & 1 & 4.3 & 1.173 & 7\\
\ 20 & 10:02:18.0 & +01:58:36.4 & 21.26 & 0.06 & 0.92 & 1 & 3.3 & 1.541 & 7\\
\ 21 & 09:59:41.4 & +01:58:45.1 & 22.11 & 0.11 & 0.98 & 1 & 4.8 & 2.502 & 7\\
\ 22 & 09:59:49.9 & +02:00:11.1 & 21.12 & 0.06 & 0.94 & 1 & 3.4 & 1.806 & 3\\
\ 23 & 10:00:17.5 & +02:00:12.7 & 21.08 & 0.07 & 0.26 & 1 & 4.9 & 0.353 & 3\\
\ 24 & 10:00:41.5 & +02:00:14.7 & 22.88 & 0.14 & 0.02 & 1 & 3.1 & - & -2\\
\ 25 & 09:59:58.9 & +02:00:24.0 & 22.16 & 0.09 & 0.97 & 1 & 3.3 & 1.033 & 7\\
\ 26 & 10:01:41.4 & +02:00:50.9 & 22.15 & 0.14 & 0.91 & 1 & 6.5 & 2.27 & 7\\
\ 27 & 10:01:08.6 & +02:00:52.6 & 20.98 & 0.05 & 0.93 & 1 & 3.1 & 2.671 & 3\\
\ 28 & 09:59:41.5 & +02:01:36.8 & 22.48 & 0.13 & 0.98 & 1 & 4.0 & 2.918 & 7\\
\ 29 & 10:01:14.9 & +02:02:08.7 & 21.33 & 0.09 & 0.63 & 1 & 7.2 & 0.971 & 3\\
\ 30 & 10:01:11.0 & +02:02:26.4 & 22.50 & 0.12 & 0.03 & 1 & 3.3 & - & -1\\
\ 31 &10:01:47.0 & +02:02:36.6 & 20.93 & 0.08 & 0.91 & 1 & 6.8 & 1.171 & 7\\
\ 32 & 10:01:20.3 & +02:03:41.2 & 19.99 & 0.06 & 0.88 & 1 & 4.4 & 0.906 & 7\\
\ 33 & 09:59:21.1 & +02:05:33.2 & 23.00 & 0.65 & 0.05 & 1 & 24.5 & - & -1\\
\ 34 & 09:59:35.5 & +02:05:38.1 & 22.25 & 0.15 & 0.98 & 1 & 6.9 & 1.91 & 7\\
\ 35 & 10:01:29.7 & +02:06:43.3 & 20.81 & 0.08 & 0.95 & 1 & 7.4 & 1.916 & 3\\
\ 36 & 10:00:25.3 & +02:07:34.7 & 21.98 & 0.14 & 0.95 & 1 & 8.0 & 0.963 & 7\\
\ 37 & 10:00:47.8 & +02:07:57.0 & 19.57 & 0.04 & 0.89 & 1 & 3.3 & 2.161 & 7\\
\ 38 & 10:00:38.0 & +02:08:22.5 & 20.60 & 0.05 & 0.92 & 1 & 3.1 & 1.825 & 7\\
\ 39 & 10:01:56.3 & +02:09:43.4 & 22.31 & 0.12 & 0.98 & 1 & 4.5 & 1.641 & 7\\
\ 40 & 10:00:20.6 & +02:12:35.0 & 21.52 & 0.09 & 0.07 & 1 & 4.5 & - & -2\\
\ 41 & 10:00:06.9 & +02:12:35.7 & 21.41 & 0.14 & 0.92 & 1 & 12.0 & 1.258 & 7\\
\ 42 & 09:59:34.9 & +02:14:22.2 & 20.91 & 0.06 & 0.93 & 1 & 4.0 & 1.734 & 7\\
\ 43 & 10:01:24.0 & +02:14:46.1 & 21.38  & 0.08 & 0.90 & 1 & 6.0 & 0.894 & 7\\
\ 44 & 10:01:48.0 & +02:14:47.2 & 20.52 & 0.11 & 0.91 & 1 & 10.1 & 0.882 & 7\\
\ 45 & 09:58:49.5 & +02:16:40.7 & 21.56 & 0.07 & 0.19 & 1 & 3.9 & 0.731 & 7\\
\ 46 & 09:59:58.5 & +02:18:05.0 & 20.06 & 0.06 & 0.88 & 1 & 4.4 & 1.789 & 3\\
\ 47 & 09:59:51.0 & +02:19:01.9 & 21.52 & 0.18 & 0.08 & 1 & 15.9 & - & -1\\
\ 48 & 10:00:55.6 & +02:21:50.3 & 21.06 & 0.08 & 0.94 & 1 & 5.9 & 1.933 & 7\\
\ 49 & 09:59:46.9 & +02:22:09.3 & 21.85 & 0.14 & 0.94 & 1 & 9.4 & 0.909 & 7\\
\ 50 & 10:02:02.8 & +02:24:34.6 & 20.84 & 0.14 & 0.92 & 1 & 13.9 & 0.987 & 7\\
\ 51 & 10:01:45.2 & +02:24:56.8 & 20.67 & 0.09 & 0.91 & 1 & 7.7 & 2.032 & 3\\
\ 52 & 10:01:18.6 & +02:27:39.0 & 20.50 & 0.08 & 0.62 & 1 & 7.4 & 1.042 & 7\\
\ 53 & 10:01:37.7 & +02:28:43.9 & 21.59 & 0.07 & 0.32 & 1 & 3.5 & 0.367 & 3\\
\ 54 & 10:01:48.9 & +02:31:40.6 & 20.99 & 0.05 & 0.91 & 1 & 3.2 & 1.935 & 7\\
\ 55 & 10:02:15.3 & +02:32:10.0 & 22.15 & 0.18 & 0.02 & 1 & 9.5 & - & -1\\
\ 56 & 09:59:08.6 & +02:33:17.2 & 22.93 & 0.15 & 0.93 & 1 & 3.1 & 1.798 & 7\\
\ 57 & 10:01:20.3 & +02:33:41.1 & 19.96 & 0.05 & 0.88 & 1 & 3.4 & 1.834 & 7\\
\ 58 & 09:59:11.1 & +02:33:50.7 & 20.84 & 0.08 & 0.91 & 1 & 5.8 & 0.704 & 7\\
\ 59 & 10:01:16.3 & +02:36:07.3 & 21.00 & 0.11 & 0.56 & 1 & 9.7 & 0.959 & 7\\
\ 60 & 10:02:23.4 & +02:37:04.5 & 21.99 & 0.14 & 0.87 & 1 & 7.8 & 1.447 & 7\\
\ 61 & 10:00:10.2 & +02:37:44.9 & 20.85 & 0.07 & 0.90 & 1 & 5.1 & 1.56 & 7\\
\ 62 & 09:59:13.9 & +02:38:44.4 & 21.69 & 0.09 & 0.90 & 1 & 5.4 & 2.082 & 7\\
\ 63 & 10:01:59.4 & +02:39:35.8 & 20.29 & 0.07 & 0.88 & 1 & 6.2 & 0.851 & 7\\
\ 64 & 10:02:01.3 & +02:40:29.7 & 21.48 & 0.11 & 0.97 & 1 & 8.0 & - & 7\\
\ 65 & 10:00:10.9 & +02:41:18.6 & 22.07 & 0.09 & 0.98 & 1 & 3.7 & 1.436 & 7\\
\ 66 & 10:02:02.2 & +02:41:57.9 & 21.74 & 0.12 & 0.35 & 1 & 8.2 & 0.794 & 7\\
\ 67 & 10:00:02.0 & +02:42:16.5 & 21.33 & 0.13 & 0.16 & 1 & 11.7 & 0.58 & 3\\
\ 68 & 10:00:14.1 & +02:43:49.2 & 21.68 & 0.10 & 0.40 & 1 & 6.9 & - & 4\\
\ 69 & 10:00:20.8 & +02:43:57.4 & 22.43 & 0.12 & 0.04 & 1 & 3.6 & - & -1\\
\ 70 & 09:58:54.3 & +02:44:13.4 & 22.89 & 0.19 & 0.02 & 1& 4.9 & - & -2\\
\ 71 & 10:01:40.4 & +02:05:06.6 & 20.58 & 0.06 & 0.03 & 2 & 4.7 & 0.425 & 3\\
\ 72 & 09:59:12.7 & +02:06:57.9 & 21.29 & 0.06 & 0.03 & 2 & 3.3 & - & -1\\
\ 73 & 10:01:41.3 & +02:10:31.5 & 21.58 & 0.12 & 0.88 & 2 & 10.0 & 0.983 & 7\\
\ 74 & 09:59:39.0 & +02:12:01.2 & 20.54 & 0.05 & 0.91 & 2 & 3.5 & 0.689 & 7\\
\ 75 & 09:59:58.5 & +02:15:30.6 & 21.37 & 0.09 & 0.04 & 2 & 7.2 & 0.659 & 3\\
\ 76 & 10:00:14.9 & +02:27:18.1 & 21.37 & 0.07 & 0.14 & 2 & 4.7 & 0.73 & 7\\
\ 77 & 10:01:48.2 & +02:31:02.8 & 21.76 & 0.33 & 0.96 & 2 & 5.7 & - & 00\\
\ 78 & 10:01:32.7 & +02:32:33.8 & 22.27 & 0.34 & 0.87 & 2 & 7.6 & - & 00\\
\ 79 & 09:59:06.7 & +02:34:02.1 & 21.76 & 0.10 & 0.79 & 2 & 6.2 & - & 6\\
\ 80 & 10:01:53.2 & +02:36:12.1 & 21.86 & 0.30 & 0.70 & 2 & 5.4 & - & 00\\
\ 81 & 10:01:32.7 & +02:41:49.8 & 22.11 & 0.30 & 0.04 & 2 & 18.4 & - & 0\\
\ 82 & 09:59:19.8 & +02:42:38.5 & 21.18 & 0.05 & 0.91 & 2 & 3.0 & 2.121 & 7\\
\ 83 & 10:01:31.9 & +02:43:24.6 & 22.75 & 0.24 & 0.85 & 2 & 8.1 & 2.048 & 7\\
\hline
\end{longtable}
\end{landscape}

\end{document}